\documentclass[superscriptaddress,preprint,showpacs,pra,longbibliography]{revtex4-1}

\usepackage{graphicx,float, placeins}
\usepackage{physics}
\usepackage{amsmath}
\usepackage{subcaption}
\usepackage{bm}
\usepackage{xcolor}

\begin{document}


\title{Dynamics of Majorana Fermions on a Quantum Computer}

\author{Yuxiao Hang\textsuperscript{1}}
\email{yhang@usc.edu}
\author{Rosa Di Felice\textsuperscript{1}}
\email{difelice@usc.edu}
\author{Aiichiro Nakano\textsuperscript{1}}
\email{anakano@usc.edu}
\author{Stephan Haas\textsuperscript{1}}%
\email{shaas@usc.edu}
 
\affiliation{\textsuperscript{1}Department of Physics and Astronomy, University of Southern California,
Los Angeles, CA 90089-0484, USA
}%

\date{\today}

\begin{abstract}
The study of quasi-particle dynamics is central to understanding non-equilibrium phenomena in quantum many-body systems. Direct simulation of such dynamics on quantum hardware has been limited by circuit depth and noise constraints. In this work, we use a recently developed constant-depth circuit algorithm to examine the real-time evolution of site-resolved magnetization in a  transverse-field Ising chain on noisy intermediate-scale quantum devices. 
By representing each spin as a pair of Majorana fermions, we identify two distinct dynamical regimes governed by the relative strength of 
spin interaction. 
Furthermore, we show how local impurities can serve as probes of Majorana modes, acting as dynamical barriers in the weak coupling regime. 
These results demonstrate that constant-depth quantum circuits provide a viable route for studying quasi-particle propagation and for probing Majorana signatures on currently available quantum processors.

\end{abstract}

\maketitle


\section{Introduction}

Since the discovery of topological insulators, the study of emergent quasi-particles has attracted great interest~\cite{PhysRevLett.95.226801,doi:10.1126/science.1148047,RevModPhys.82.3045,RevModPhys.83.1057}. Among these, Majorana fermions \cite{Elliott_RMP2015_majorana} have received particular attention due to their non-Abelian statistics and potential applications in topological quantum computation. 
The use of Majorana fermions as a basis to develop a quantum computational platform was first proposed by Kitaev~\cite{Kitaev_2001}. Subsequently,  
Majorana fermions have been investigated both theoretically and experimentally, including studies of their dynamics under perturbations~\cite{PhysRevB.85.020502} and experimental realizations in engineered nanowire systems~\cite{doi:10.1126/science.1148047}. However, despite their promise, their technological use remains challenging. 
The real time evolution dynamics of Majorana fermions — particularly their behavior in driven or perturbed systems — remains relatively unexplored.

A quantum quench entails the preparation of a many-body quantum state in the ground state of an initial Hamiltonian, $H_i$, followed by the sudden change to a new Hamiltonian, $H_f$, which determines the system's time evolution. This protocol has been applied to the study of the entanglement dynamics of many-body systems, such as the transverse field Ising model (TFIM) \cite{Pasquale_Calabrese_2004,Calabrese_2005,10.21468/SciPostPhysLectNotes.20,10.21468/SciPostPhys.5.4.033,PhysRevA.69.053616}. 
Using traditional computers, these investigations have revealed that isolated many-body systems, even under purely unitary evolution, can exhibit emergent thermalization, universal entanglement growth, and dynamical phase transitions governed by underlying conservation laws and the structure of excitations.

The emergence of quantum computing devices offers a new route for studying the dynamics of quantum many-body systems. In particular, quantum spin chains have become a natural target for simulation on noisy intermediate-scale quantum (NISQ) devices, because of the formal similarity between qubits and spin-$\frac{1}{2}$ sites~\cite{PhysRevB.110.075116,PhysRevResearch.5.013183,Bassman_MT2022,
Lancaster_AJP2025,Gandon_PRXQ2025,LOTSTEDT202433,Monroe_RMP2021,Fauseweh_NatureComm2024}. Most studies to date have focused on quantities such as on-site magnetization or ground state energies \cite{Mercado_PRB2024}. There has been little investigation into the dynamics of emergent quasi-particles 
\cite{Balents_PRA2025} like Majorana fermions on such platforms.

The TFIM provides a paradigmatic example of a spin chain that supports Majorana fermions \cite{Narozhny_SciRep2017,Seiberg_ScipostPhys2024,PhysRevLett.132.210401,hn66-j8pt}. Through the Jordan-Wigner transformation, each spin site can be mapped to a pair of Majorana operators \cite{Seiberg_ScipostPhys2024}. 
In the TFIM, the regime $J > h$ corresponds to the ordered, topological phase. In this phase, unpaired Majorana modes appear at the two ends of an open chain, leading to a twofold degeneracy. Importantly, this degeneracy is not limited to the ground state but extends throughout the entire many-body spectrum, a phenomenon known as a \emph{Strong Zero Mode}\cite{Kemp_2017,Kitaev_2001}. The associated edge Majorana operators, $\Gamma_L$ and $\Gamma_R$, commute with the Hamiltonian - up to exponentially small corrections in system size - and they map every eigenstate to a nearly degenerate partner of opposite fermion parity. 
\cite{Fendley_2012,Kitaev_2001}. 
As a consequence, the Majorana zero modes persist at the boundary even after quenches that populate highly excited states. Furthermore, the model's quadratic Fermionic Hamiltonian 
allows for efficient simulation of its dynamics because of recent advances in quantum algorithms, i.e.,  the Constant Depth Circuit (CDC) approach \cite{Bassman_MT2022,PhysRevB.110.075116}. This method enables high-fidelity simulations of long-time dynamics of free fermion systems on NISQ-era devices \cite{Preskill2018quantumcomputingin,Mercado_PRB2024}.

\begin{equation}
    H = J \sum_{n=1}^{N} \sigma^x_n \sigma^x_{n+1} + h \sum_{n=1}^{N+1} \sigma^z_n
    \label{eq:tfim_hamiltonian_original}
\end{equation}

In this work, we apply the Constant Depth Circuit algorithm to study the long-time evolution of Majorana fermions in the TFIM using an IBM quantum computer. By simulating the model on a 7-site chain, we identify distinct dynamical regimes tuned by the relative strength of the inter-site coupling versus the transverse field. In addition, we introduce localized impurities into the chain and investigate their influence on the dynamics of Majorana fermions.

The study of impurities in quantum chains  has a long and rich history in condensed matter physics~\cite{RevModPhys.55.331,doi:10.1126/science.1222360,10.21468/SciPostPhysCore.8.1.002}. In the context of the TFIM, analytical work has examined the effect of impurity-localized modes~\cite{MULLER2016482}. Here, we demonstrate how a local impurity in the bulk can modify the evolution of the Majorana modes, alter transport properties, and effectively split the chain.

This paper is organized as follows. In Sec.~II, we introduce the TFIM Hamiltonian and its mapping onto Majorana fermions. In Sec.~III, we present NISQ simulation results, including the identification of different dynamical regimes, impurity-induced phenomena, such as evidence for mid-chain Majorana modes. We conclude with a summary of our findings and an outlook for future work. In the Appendix, we provide technical details regarding the derivations of the Majorana representation and further computational details.

\section{The Transverse Field Ising Model in Majorana Representation}

\begin{figure}
\centering
\subcaptionbox{\protect\label{a}}
{\includegraphics[width=0.90\linewidth,height=0.15\linewidth]{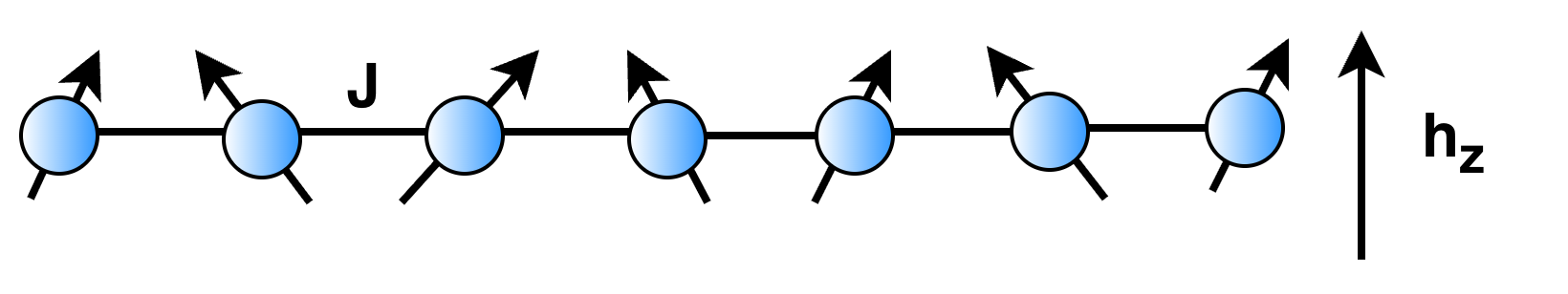}}\hfill
\centering
\subcaptionbox{\protect\label{b}}
{\includegraphics[width=0.95\linewidth,height=0.07\linewidth]{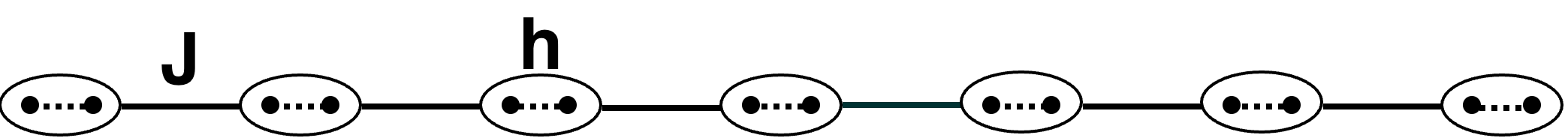}}\hfill
\subcaptionbox{\protect\label{c}}
{\includegraphics[width=0.95\linewidth,height=0.07\linewidth]{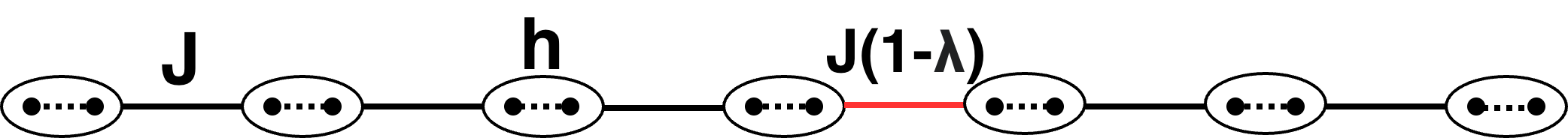}}\hfill
\caption{Illustration of the (a) transverse field Ising model on an open ended 7 qubit chain and (b) its representation using Majorana fermions with no impurities. (c) Schematic illustration of a 7-qubit chain with an impurity coupling located between site 4 and site 5. The inter-cell coupling in (a-c) is illustrated using solid lines, while the intra-cell coupling in the fermionic model in (b,c) is illustrated using dashed lines. The spin site in (a) is represented using a sphere with arrow. The fermionic site in (b,c) is represented by a dot. The red line in (c) indicates an impurity-like inter-cell coupling strength.}
\label{figure:model}
\end{figure}

We consider the TFIM on a one-dimensional spin-$\frac{1}{2}$ chain with length $N+1$ 
and open boundary conditions. In addition, we place an impurity on the central site. The Hamiltonian, in spin language, is then given by
\begin{equation}
    H = J \sum_{n=1}^{N} \sigma^x_n \sigma^x_{n+1} + h \sum_{\substack{n=1 \\ n \neq \frac{N}{2}+1}}^{N+1} \sigma^z_n + h(1 - \lambda)\, \sigma^z_{\frac{N}{2}+1},
    \label{eq:tfim_hamiltonian}
\end{equation}
where $\sigma^\alpha_n$ ($\alpha = x, z$) are Pauli matrices acting on the $n$-th spin site, $J$ denotes the nearest-neighbor exchange coupling, and $h$ is the transverse magnetic field magnitude. The impurity enters via a local magnetic field term with tuning parameter $\lambda $. Throughout the numerical study, we focus on a chain of length $N+1=7$ (Fig. \ref{figure:model}(a)).

Using the Jordan-Wigner transformation, the spin operators can be mapped onto spinless fermions, whereby each spin site is represented by two Majorana fermion operators. The resulting Hamiltonian, expressed in terms of Majorana modes, takes the form
\begin{align}
    H = & J \sum_{n=1}^{N} i \gamma_{2n} \gamma_{2n+1} + h \sum_{\substack{n=1 \\ n \neq \frac{N}{2}+1}}^{N+1} 
    + ih(1-\lambda)\gamma_{[2(\frac{N}{2}+1)-1]}\gamma_{[2(\frac{N}{2}+1)]},
\end{align}
where $\gamma_j$ are Majorana operators satisfying $\gamma_j^\dagger = \gamma_j$ and $\{ \gamma_j, \gamma_k \} = 2\delta_{jk}$. The system thus maps onto a tight-binding model of $2(N+1)$ 
Majorana fermions, as illustrated in Fig. \ref{figure:model}(b). Each spin site corresponds to a pair of Majorana modes, with $h$ describing the intra-site coupling between two Majorana modes belonging to the same spin site (on-site spin bias), and $J$ describing the inter-site (nearest-neighbor) coupling between two Majorana modes belonging to two adjacent spin sites. In this Majorana representation, 
the Majorana operators at both edges of the chain ($\gamma_1$ and $\gamma_{14}$) are excluded from inter-spin-site coupling, and the impurity term  acts on the central link. 

In this model, in the absence of impurities, the relative strengths of \( J \) and \( h \) 
control the phase space. 
In the regime \( J > h \), the system enters a \textit{topologically nontrivial} phase, characterized by two unpaired Majorana zero modes localized at the edges of the open chain \cite{condmat6010011}. These edge modes evolve differently from the bulk modes and are robust signatures of topological protection. In contrast, when \( h > J \), the system is in a \textit{topologically trivial} (paramagnetic) phase, where the transverse field dominates. In this regime, the Majorana fermions are strongly hybridized within each site, and edge and bulk modes behave similarly, with dynamics governed primarily by the local field \( h \), leading to largely decoupled and uniform evolution across the chain.

The Majorana Zero Modes (MZMs) are  represented by two operators, $\Gamma_L$ and $\Gamma_R$,  expressed as
\begin{align}
    &\Gamma_L = \sum_{n=1}^{N+1}(\frac{h}{J})^{n}\gamma_{2n-1} \nonumber \\ 
    &\Gamma_R = \sum_{n=1}^{N+1}(\frac{h}{J})^{N-n+1}\gamma_{2n}
\end{align}

In the $h \ll J$ limit the zero-mode wavefunctions decay as $(h/J)^n$, so the Majorana operators are exponentially localized at the two ends of an open chain. Under the inverse Jordan--Wigner transformation, they map to boundary spin-flip operators (up to the global $Z_2$ string) that connect the two degenerate Ising vacua. Consequently, local edge spins have a finite overlap with the zero modes, leading to long-lived boundary autocorrelations and making the Majorana zero modes directly visible in the dynamics of edge spins, as we will observe below.

\section{Time Evolution via Constant Depth Circuits}

Traditionally, quantum time evolution is implemented using the Trotter-Suzuki decomposition \cite{TrotterSuzuki1,TrotterSuzuki2,TrotterSuzuki3}, which approximates the time-evolution operator at discretized time $n\Delta t$ via
\begin{equation}
    U(n\Delta t) \approx \prod_{\tau=1}^n \prod_{l} e^{-iH_l(t_\tau)\Delta t} + O(\Delta t),
    \label{eq:trotter}
\end{equation}
where $U(t)=\mathcal{T}\exp{\{-i\int_0^tdsH(s)}\}$ gives the transformation of the system wave function from the initial function $\psi_0$ to the function at time $t$ $\psi(t)$, $\psi(t)=U(t)\psi_0$; $\mathcal{T}$ indicates the time-order operator. Note that Eq. \ref{eq:trotter} can be applied to both time-dependent and time-independent Hamiltonians.
Here, the $H_l$ are easy-to-diagonalize terms in a decomposition of the Hamiltonian, and the evolution time is discretized into $n$ time intervals of duration $\Delta t$ during which the system Hamiltonian is approximately constant. 
This approach involves a trade-off between the Trotter step size $\Delta t$ and the circuit depth. A small $\Delta t$ reduces the Trotter error, $O(\Delta t)$, but increases the total number of steps needed to reach a given final time, thus increasing circuit depth—typically beyond the coherence limits of near-term quantum devices. The Trotter error can be improved to $O(\Delta t^3)$, but the tradeoff between accuracy and feasibility remains an issue. 

For \textit{quadratic Hamiltonians}, such as the TFIM, this problem can be circumvented using the \textit{constant depth circuit} (CDC) algorithm. As we discuss here, CDCs allow time evolution to be implemented with a depth that remains constant regardless of the total simulated time, thus enabling arbitrarily small $\Delta t$ without increasing circuit depth and thereby avoiding the Trotter error entirely.

The CDC construction proceeds as follows:
\begin{enumerate}
    \item Prepare a quantum circuit with $N$ qubits, where $N$ equals the number of spins in the target system.
    \item Construct \textit{matchgates} (shown as G in Fig. \ref{figure:Final derivation} in Appendix C), which  are the building blocks of the constant depth circuit for the desired Hamiltonian. At each time step, these are two-qubit gates acting on neighboring qubits in a brickwork pattern\cite{Bassman_Oftelie_2022}. The \textit{matchgates} depend on parameters that are unknown and must be determined. 
    \item Numerically optimize the parameters of these \textit{matchgates} for each target time to reproduce the desired time-evolution unitary. The gate structure and the corresponding parameter matrix for the TFIM was derived in \cite{Bassman_Oftelie_2022}. For the TFIM, the corresponding parameters for each G appear in the form
    \begin{equation}
        G = \begin{bmatrix}
            e^{-i(\theta_0+\theta_3)}\cos(\frac{\theta_1-\theta_2}{2}) & 0 & 0 & -ie^{i(\theta_0-\theta_3)}\sin(\frac{\theta_1-\theta_2}{2}) \\ 0
             & \cos(\frac{\theta_1+\theta_2}{2}) & -i\sin(\frac{\theta_1+\theta_2}{2}) & 0 \\ 0 & -i\sin(\frac{\theta_1+\theta_2}{2}) & \cos(\frac{\theta_1+\theta_2}{2}) & 0 \\ -ie^{i(\theta_0-\theta_3)}\sin(\frac{\theta_1-\theta_2}{2}) & 0 & 0 & e^{i(\theta_0+\theta_3)}\cos(\frac{\theta_1-\theta_2}{2})
             
        \end{bmatrix}
    \end{equation}
    \item Implement the circuit on IBM and run the simulation.
\end{enumerate}

Quadratic Hamiltonians are amenable to this approach due to two key algebraic properties of \textit{matchgates}: composability and mirroring. 

\textbf{Composability:} Two successive \textit{matchgates} $G(A_1, B_1)$ and $G(A_2, B_2)$ can be composed into a single matchgate, 
$
    G(A_3, B_3) = G(A_1, B_1) G(A_2, B_2),
    $
with $A_3 = A_1 A_2$ and $B_3 = B_1 B_2$.

\textbf{Mirroring:} Any sequence of three alternating \textit{matchgates} can be rewritten in reverse order. That is, for any $G_1, G_2, G_3$, there exist \textit{matchgates} $G_4, G_5, G_6$ such that
$
    (G_1 \otimes I)(I \otimes G_2)(G_3 \otimes I) = (I \otimes G_4)(G_5 \otimes I)(I \otimes G_6).
$

These properties allow us to reduce circuit depth and rearrange the gates to a symmetric form. For the 7-qubit TFIM system, we leverage these properties to construct a mirrored, constant-depth evolution circuit (details in Appendix~\ref{Appendix:circuit derivation}), following the method developed in Ref. \cite{Bassman_Oftelie_2022}. The resulting circuit structure is shown in Fig.~\ref{figure:Final derivation}(c). \\


\section{Results}

\subsection{Benchmarking the Constant Depth Circuit algorithm}

\begin{figure}
    \centering
    \subcaptionbox{\protect\label{a}}{\includegraphics[width=0.48\linewidth,height=0.42\linewidth]{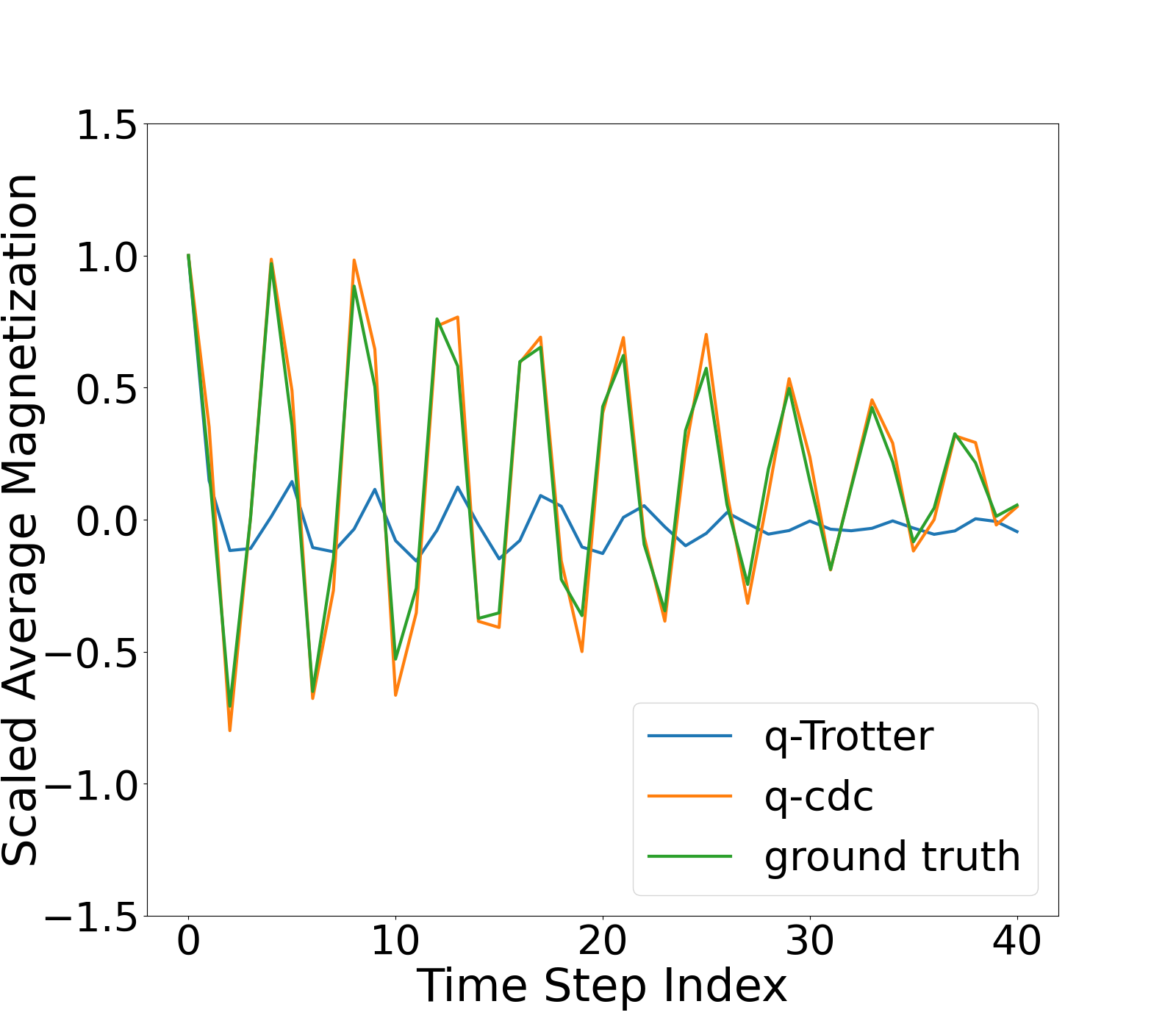}}\hfill
    \subcaptionbox{\protect\label{b}}
    {\includegraphics[width=0.48\linewidth,height=0.42\linewidth]{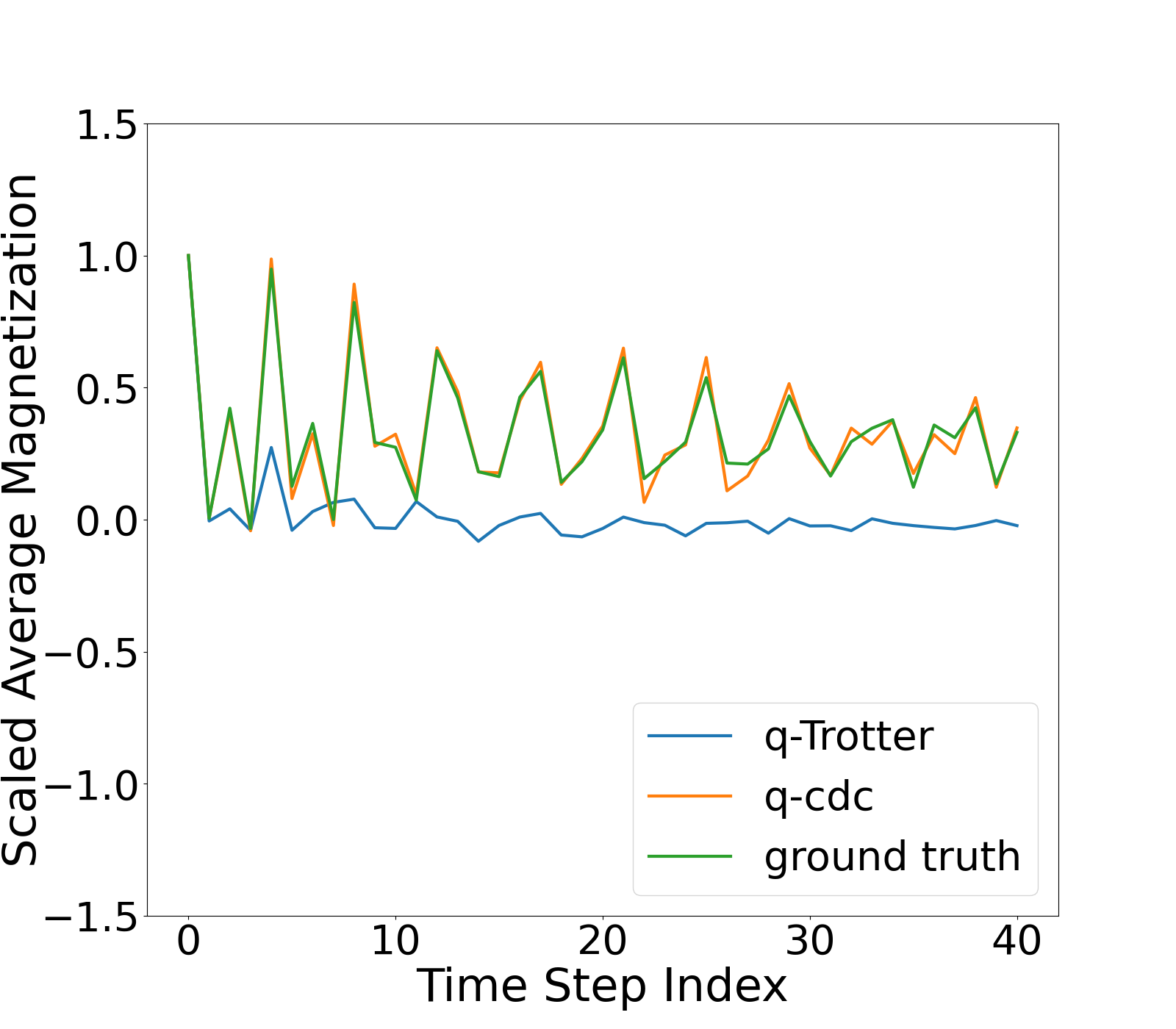}}\hfill
    \caption{Total scaled magnetization (magnetization divided by the initial magnetization), averaged over the spins, as a function of time for the 
    CDC on \texttt{ibm\_kyiv} (red), the Trotter circuit on \texttt{ibm\_kyiv} (blue) and the Trotter circuit on a noiseless simulator (green), for two different parameter regimes in Eq. \ref{eq:tfim_hamiltonian}. 
(a) Field-dominated regime with (\( J = 0.5 \) eV) and \( h = 10.0 \) eV ). 
(b) Coupling-dominated regime with (\( J = 10.0 \) eV  and \( h = 0.5 \) eV ). The duration of each time step is 0.05 fs, the total evolution time is 2 fs.
}
\label{figure:assessment_cdc}
\end{figure}

To validate the accuracy of the CDC for the Hamiltonian under study, we compared its performance against Trotter circuit execution on a noiseless simulator (ground truth) and on quantum hardware. 
%
We computed the system-averaged magnetization over 8192 measurement shots at each time step, normalized by the initial magnetization. Throughout the paper, we have chosen Hamiltonian parameters such that each time step corresponds to 0.05 fs, thus matching typical experimental conditions. $\hbar$ is chosen to be in units of eV/fs, thus making eV the units of the coupling and the magnetic field.
The simulations were performed on the IBM \texttt{ibm\_kyiv} quantum processor.
We explored different Hamiltonian parameter regimes—one dominated by the magnetic field and the other one by the interaction strength. In both cases, the CD circuit yielded results in excellent agreement with the ground truth signal, while the Trotter circuit execution on quantum hardware dramatically failed. Thus, we conclude that CDCs for TFIM are robust in noisy hardware environment, and we can use them to explore the dynamics of Majorana fermions in the wider phase space for longer times. The results of this assessment are presented in Fig. \ref{figure:assessment_cdc}. 
%

\subsection{Site resolved magnetization dynamics: quench from fully polarized initial state}

As mentioned above, the dynamics of the system differ markedly depending on the relative strength of the coupling \( J \) and the transverse field \( h \). In the regime where \( J > h \), the two Majorana fermions at the chain edges are expected to evolve differently from those in the bulk, due to the topological nature of the edge modes. In contrast, when \( h > J \), the dynamics are dominated by the local magnetic field, and the Majorana fermions evolve uniformly . 

To demonstrate the robustness of the MZMs in the \( J > h \) regime, we initialize the spins in a fully polarized configuration, with all spins aligned in the \( +z \)-direction, which is the ground state when h$\to-\infty$. We then directly measure the site-resolved magnetization along the \( z \)-direction, since the \( x \)-direction interaction term dominates their evolution. 
In the \( h > J \) regime, however, initializing in a fully polarized \( z \)-state would lead to negligible observables. Therefore, we instead prepare an initial state with all spins aligned in the \( +x \)-direction by applying Hadamard gates to each qubit. To measure the time evolution of the \( x \)-magnetization, we then apply Hadamard gates again and measure in the \( z \)-basis.

\begin{figure}
\centering

\subcaptionbox{\protect\label{a}}{\includegraphics[width=0.43\linewidth]{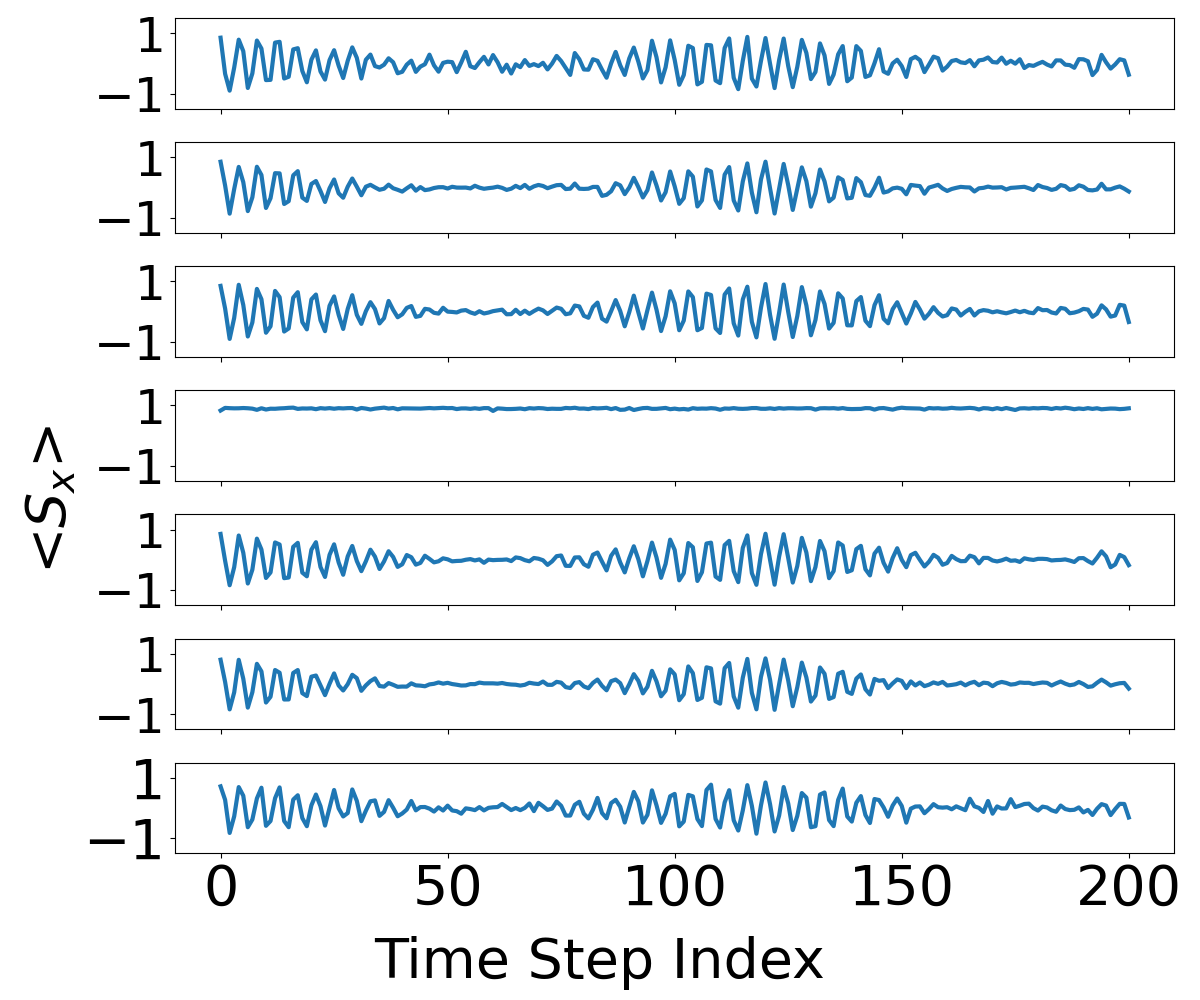}}\hfill
\subcaptionbox{\protect\label{b}}
{\includegraphics[width=0.43\linewidth]{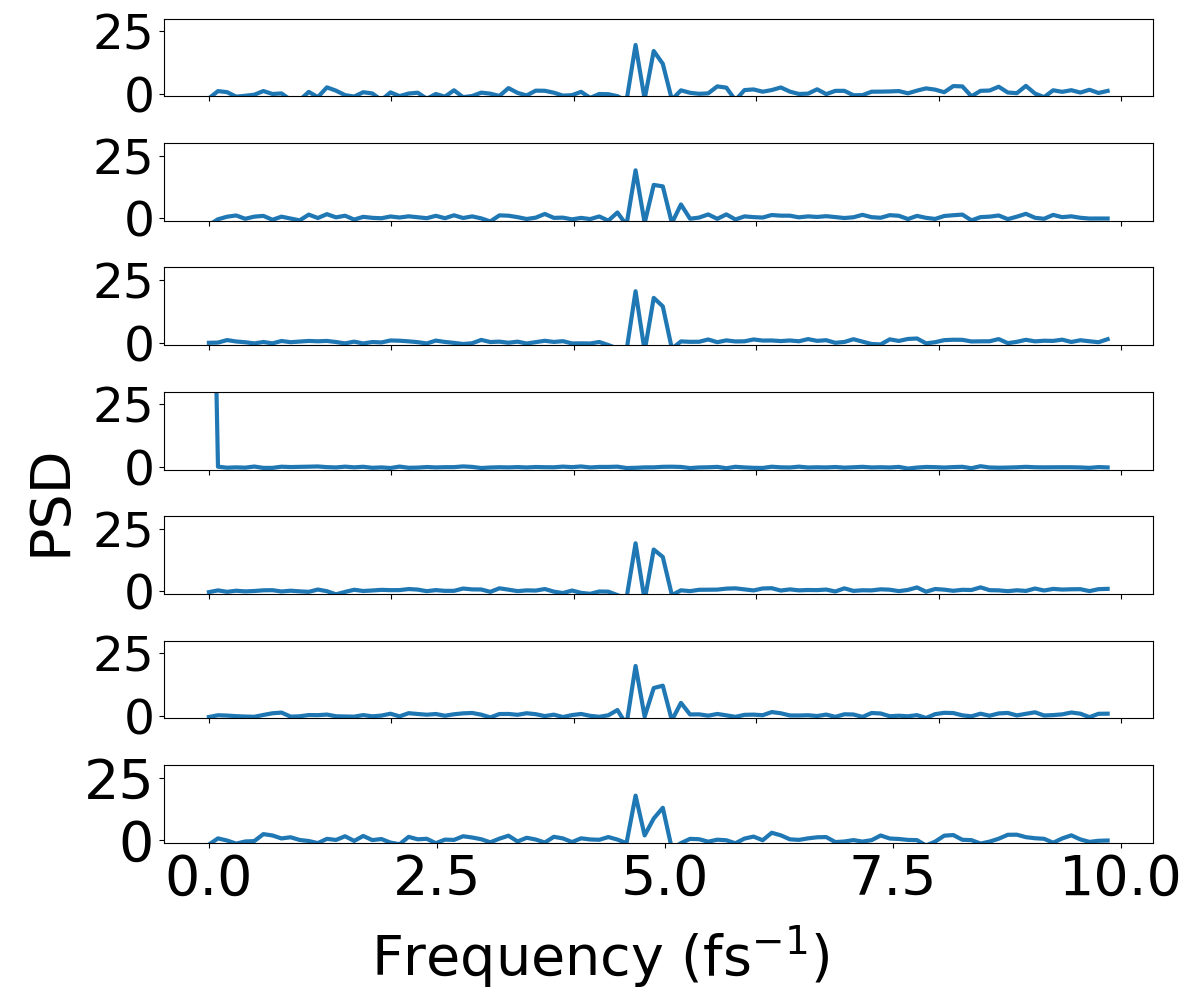}}\hfill
\subcaptionbox{\protect\label{c}}
{\includegraphics[width=0.43\linewidth]{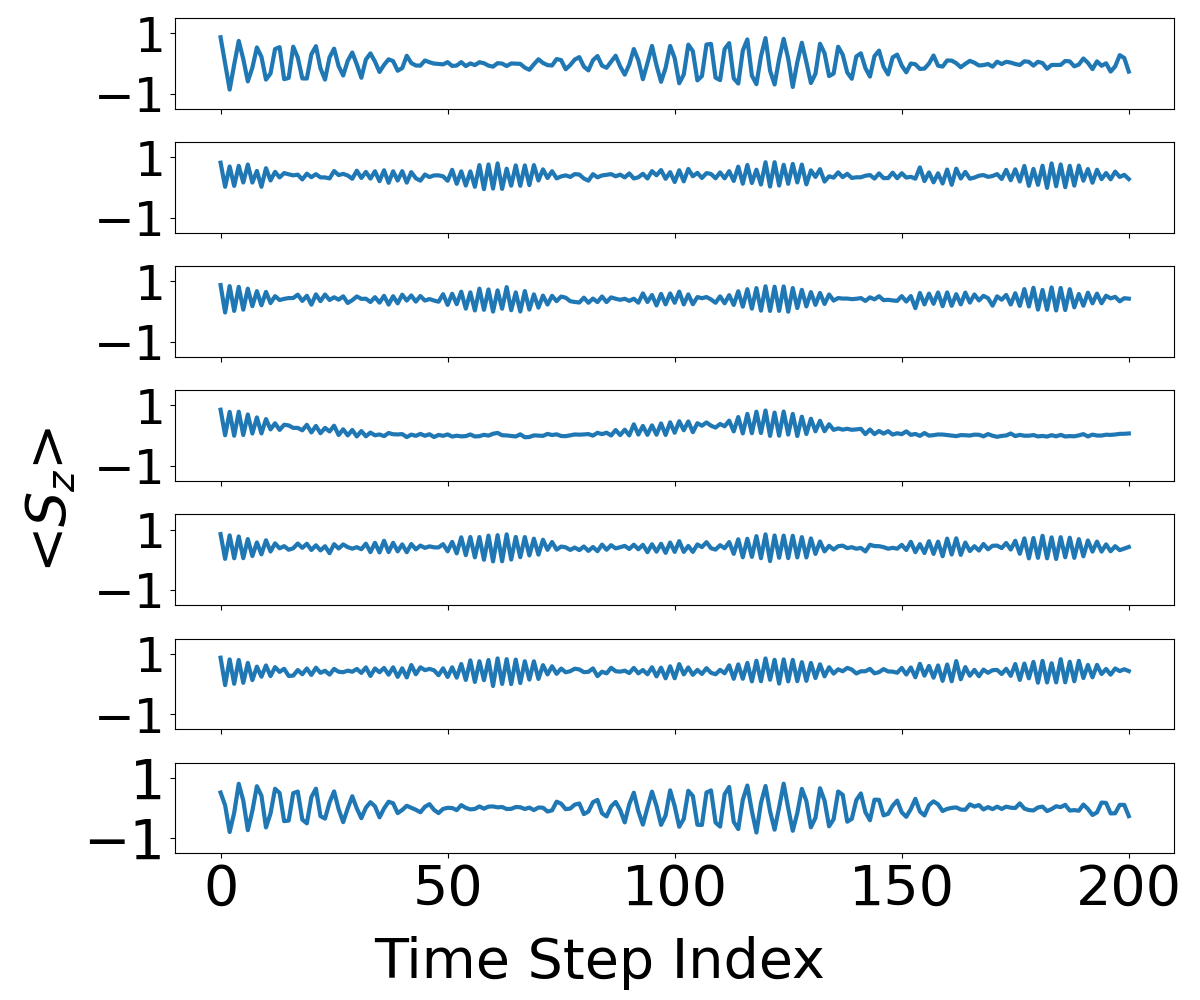}}\hfill
\subcaptionbox{\protect\label{d}}
{\includegraphics[width=0.43\linewidth]{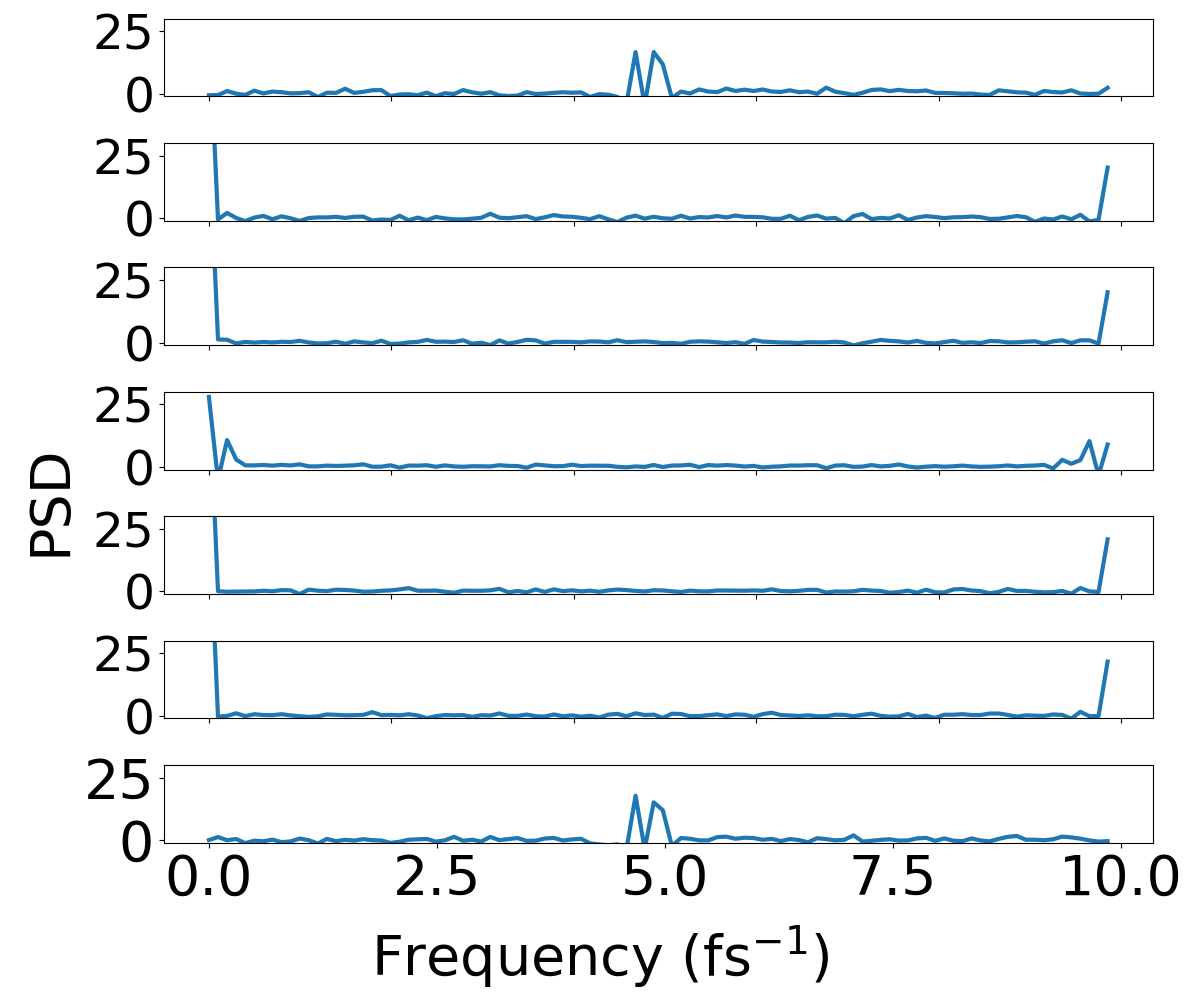}}\hfill

\caption{Time evolution of qubit magnetizations and their corresponding Fourier transforms in power spectral density (PSD). The system is initialized in fully polarized states along the x- and z-directions respectively.
(a) \( x \)-component of the site-resolved magnetizations for \( J = 0.5\) eV , \( h = 10.0 \) eV , and \( \lambda = 1.0 \) (local magnetic field at the central site is changed to  \( h \to (1-\lambda) h \)). Each sub-figure from top to bottom represents individual qubits, from qubit one to qubit seven;  
(b) Fourier transform of the data in (a);  
(c) \( z \)-component of the site-resolved z-magnetizations for \( J = 10.0 \) eV , \( h = 0.5 \) eV , and \( \lambda = 1.0 \);  
(d) Fourier transform of the data in (c). The time step is $\Delta t=0.05 ~\text{fs}$, the total evolution time is $10.0~\text{fs}$.}
\label{figure:results_extreme_parameters}
\end{figure}

Figs. \ref{figure:results_extreme_parameters}(a) and \ref{figure:results_extreme_parameters}(c) illustrate the real-time magnetization dynamics in both regimes, adding a bulk impurity at the center of the chain. Specifically, we have used \( J/h = 0.05 \) for the field-dominated regime and \( J/h = 20.0 \) for the coupling-dominated regime. Figs. \ref{figure:results_extreme_parameters}(a) and \ref{figure:results_extreme_parameters}(c) show the Fourier transformed signals, respectively. In the \( h > J \) regime (Figs. \ref{figure:results_extreme_parameters}(a) and \ref{figure:results_extreme_parameters}(b)), all non-impurity sites exhibit nearly identical time evolution (Fig. \ref{figure:results_extreme_parameters}(a)), consistent with field-dominated behavior. The impurity site, where the effective local field is set to zero, exhibits constant magnetization due to the absence of a local driving term. In the \( J > h \) regime (Fig. \ref{figure:results_extreme_parameters}(c)), the edge sites display behavior distinct from the bulk, consistent with the presence of localized Majorana edge modes. The impurity site also exhibits deviations from the bulk, though less prominently than in the field-dominated regime, as the impurity-induced variation in \( h \) is small compared to the dominant \( J \).

To further distinguish the different behavior of edge and bulk sites across different regimes, we analyze the Fourier spectra of the time-evolved magnetization signals.  The resulting site resolved dynamics, shown in Figs. \ref{figure:results_extreme_parameters}(b) and \ref{figure:results_extreme_parameters}(d), further illuminate the distinctions in dynamical behavior across regimes.
In the \( J < h \) regime, the intra-cell coupling dominates and on-site Majorana Fermions pair up. As shown in (Fig. \ref{figure:results_extreme_parameters}(b)), the edge and bulk sites exhibit similar dynamics, evidenced by nearly identical peak frequencies and intensities in Fourier space. This indicates that the evolution is dominated by the local magnetic field, and that all sites—including the edges—evolve uniformly, except the central impurity.

In contrast, in the \( J > h \) regime (Fig. \ref{figure:results_extreme_parameters}(d)), the peak frequency of the edge sites differs noticeably from that of the bulk, where the edge spins evolve with much slower frequencies compared to the bulk sites, reflecting the influence of localized Majorana modes at the system boundaries. To quantify this difference, we computed
the cumulative density(CSD)\cite{koopmans1995spectral},  defined as

\begin{equation}
    CSD{(\omega)} = \frac{\int_{0}^{\omega}PSD(\omega')d\omega'}{\int_{0}^{\infty}PSD(\omega')d\omega'}.
    \label{eq:csd}
\end{equation}

\begin{figure}
    \centering
    \subcaptionbox{\protect\label{a}}
    {\includegraphics[width=0.43\linewidth]{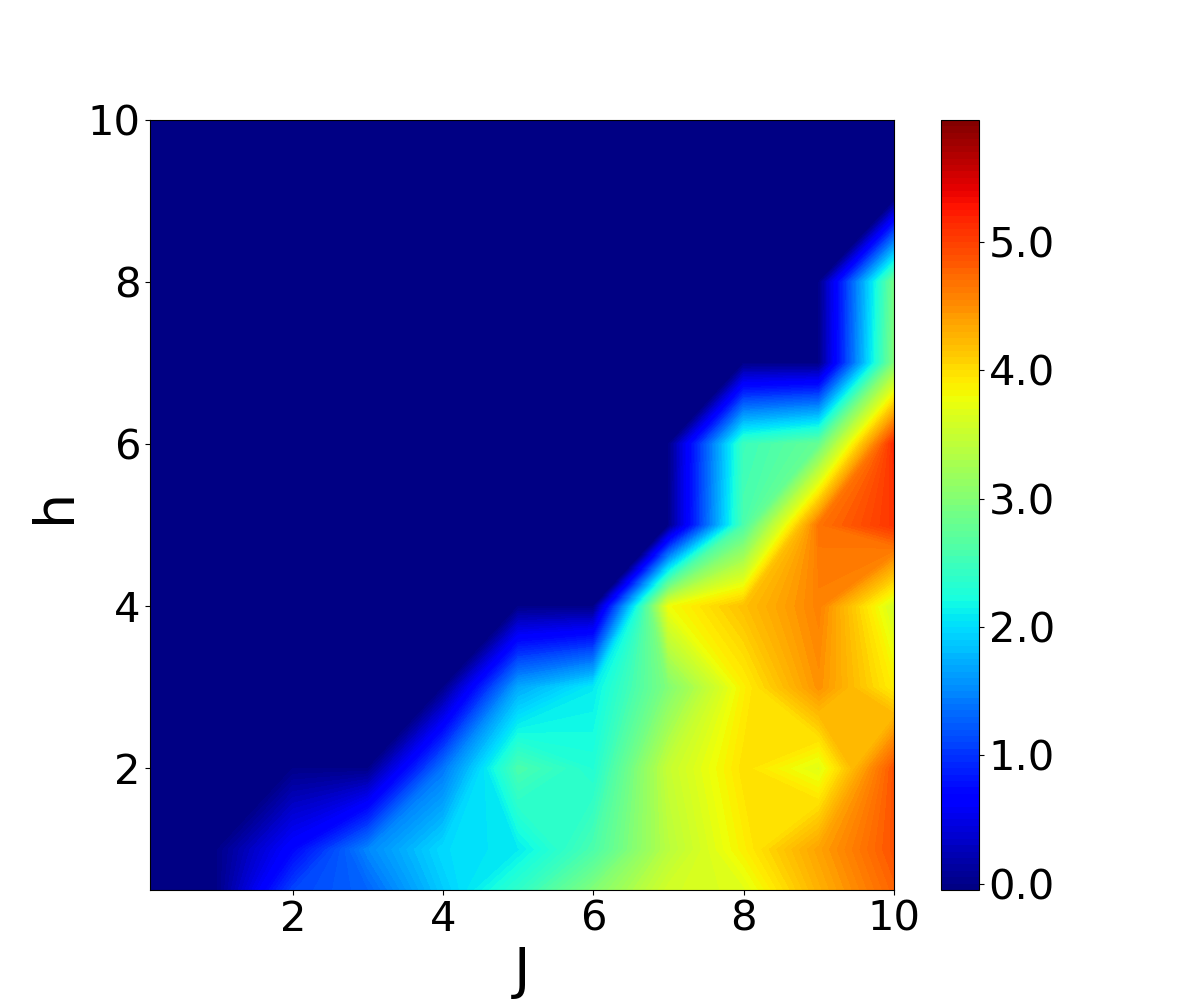}}
    \subcaptionbox{\protect\label{b}}
    {\includegraphics[width=0.43\linewidth]{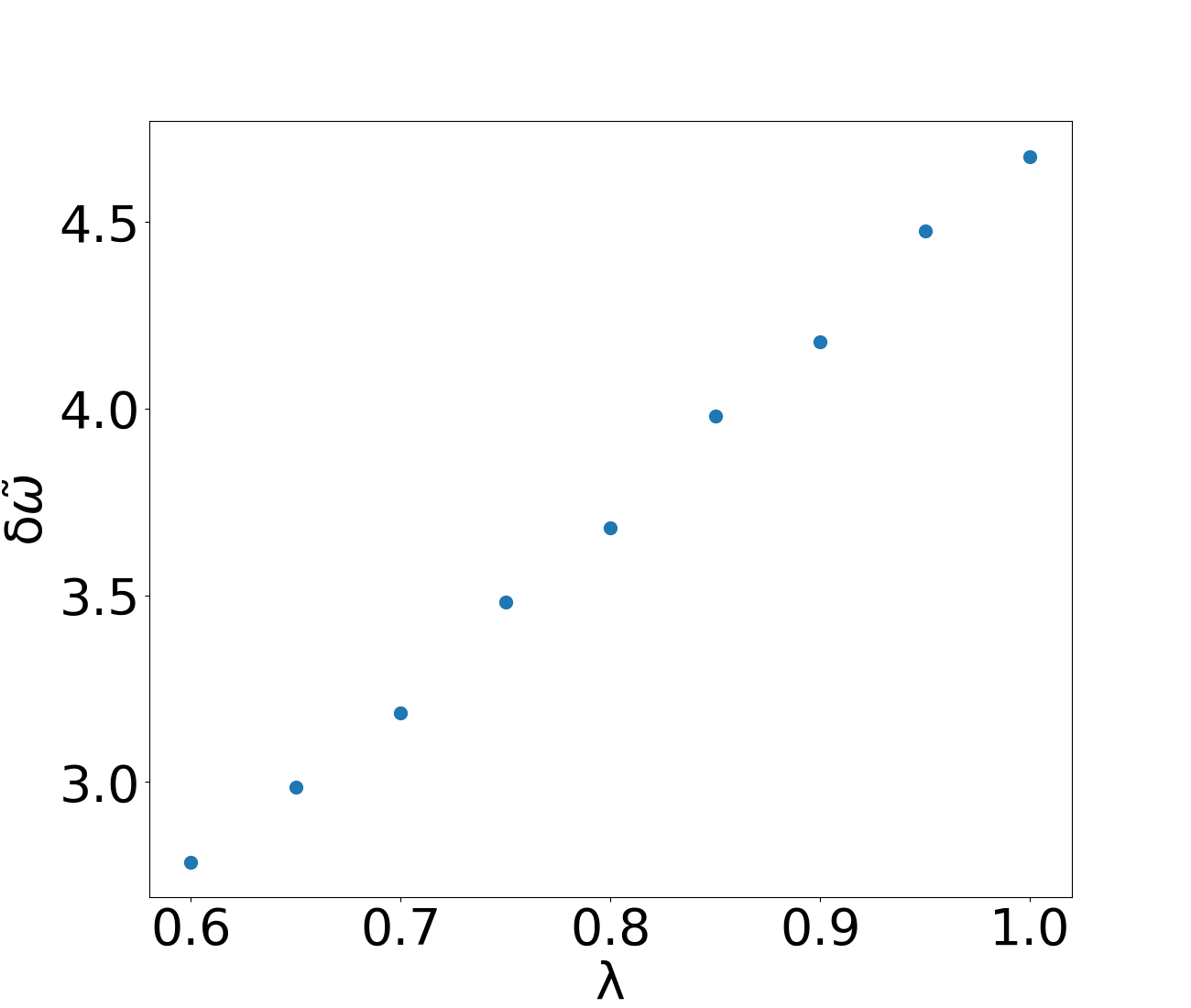}}\hfill
    \caption{(a) Heat map showing $\delta{\tilde{\omega}}$  between one of the edge sites (site 1) and a bulk site (site 2), extracted from the Fourier transform of the time-evolved site-resolved magnetization for a wide range of model parameters. This way, a phase diagram can be constructed experimentally by monitoring the system dynamics. $\tilde{\omega}$ is defined in the text after Equation \ref{eq:csd}. 
    (b) $\delta{\tilde{\omega}}$ between the impurity site ($i=4$) and a non-impurity site ($i=1$) in trivial regime ($J=0.5$ eV, $h=10.0$ eV). $\lambda$ is the strength of the local impurity at the central site which changes locally \( h \to (1-\lambda) h \), which is demonstrated in equation (3)}
    \label{figure:heatmap}
\end{figure}

Here, we focus on the frequency $\tilde{\omega}$ that corresponds to $CSD{(\tilde{\omega})}=0.5$, which means that 50 percent of spectral density is accumulated below this frequency. After extracting $\tilde{\omega}$ from the data, we calculate the difference between $\tilde{\omega}$ for bulk sites vs. edge sites, denoted as $\delta{\tilde{\omega}}$. In Fig. \ref{figure:heatmap}(a), we then plot a heat map of $\delta{\tilde{\omega}}$ over a broad range of parameters J and h.

 As the inter-site coupling \( J \) increases while keeping the on-site field \( h \) constant, namely along horizontal lines in Fig. \ref{figure:heatmap}(a), the separation between the edge and bulk frequencies increases. This effect is striking for small values of \( h \) and becomes weaker at larger values of \( h \), as one moves from bottom to top in the map. Indeed, for large values of \( h \) the edge-bulk separation disappears, as the dynamics is dominated by identical on-site time evolution. Although this behavior is expected, it is remarkable to obtain it with a noisy quantum computer over a wide parameter space and for statistically significant evolution times. 
 

\subsection{Site resolved magnetization dynamics: quench from initially staggered states}

In the previous subsection, we examined the quench dynamics, starting from an initial fully polarized state. In this section, we study the quench dynamics, starting from staggered states. Similar to the previous subsection, we analyze the two distinct parameter regimes and again find robustness of the MZM in the topological regime.

\begin{figure}
\centering

\subcaptionbox{\protect\label{a}}{\includegraphics[width=0.43\linewidth]{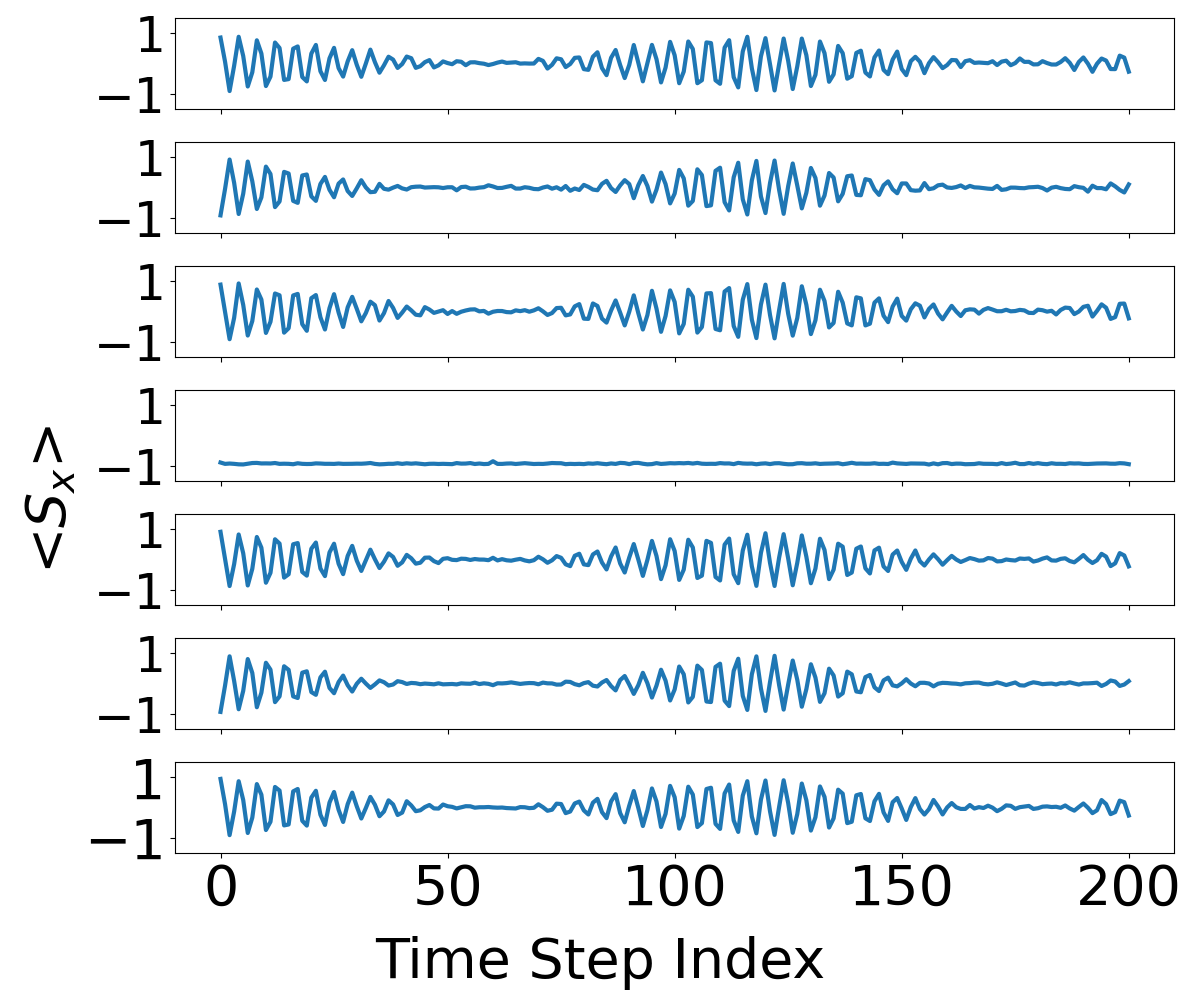}}\hfill
\subcaptionbox{\protect\label{b}}
{\includegraphics[width=0.43\linewidth]{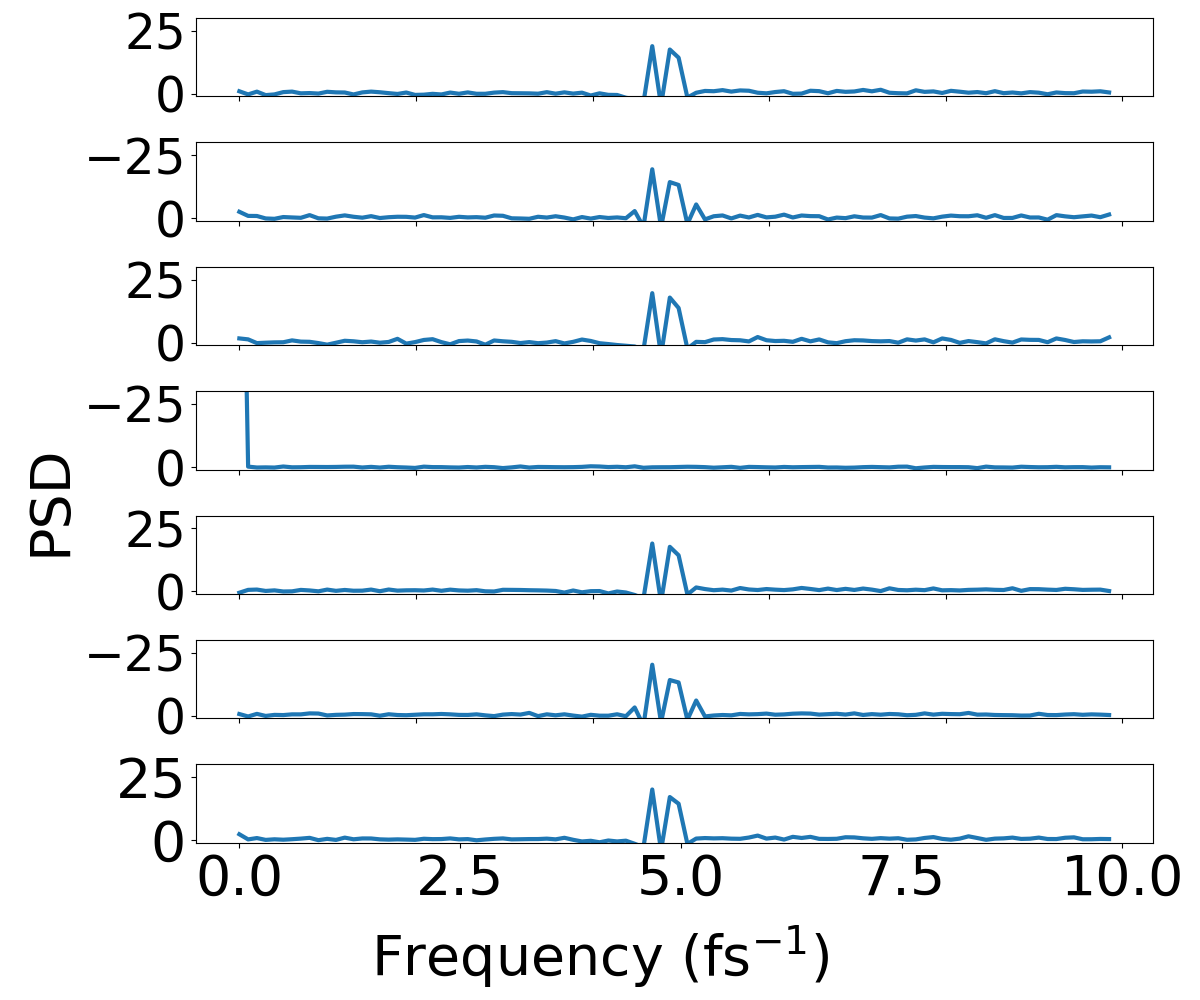}}\hfill
\subcaptionbox{\protect\label{c}}
{\includegraphics[width=0.43\linewidth]{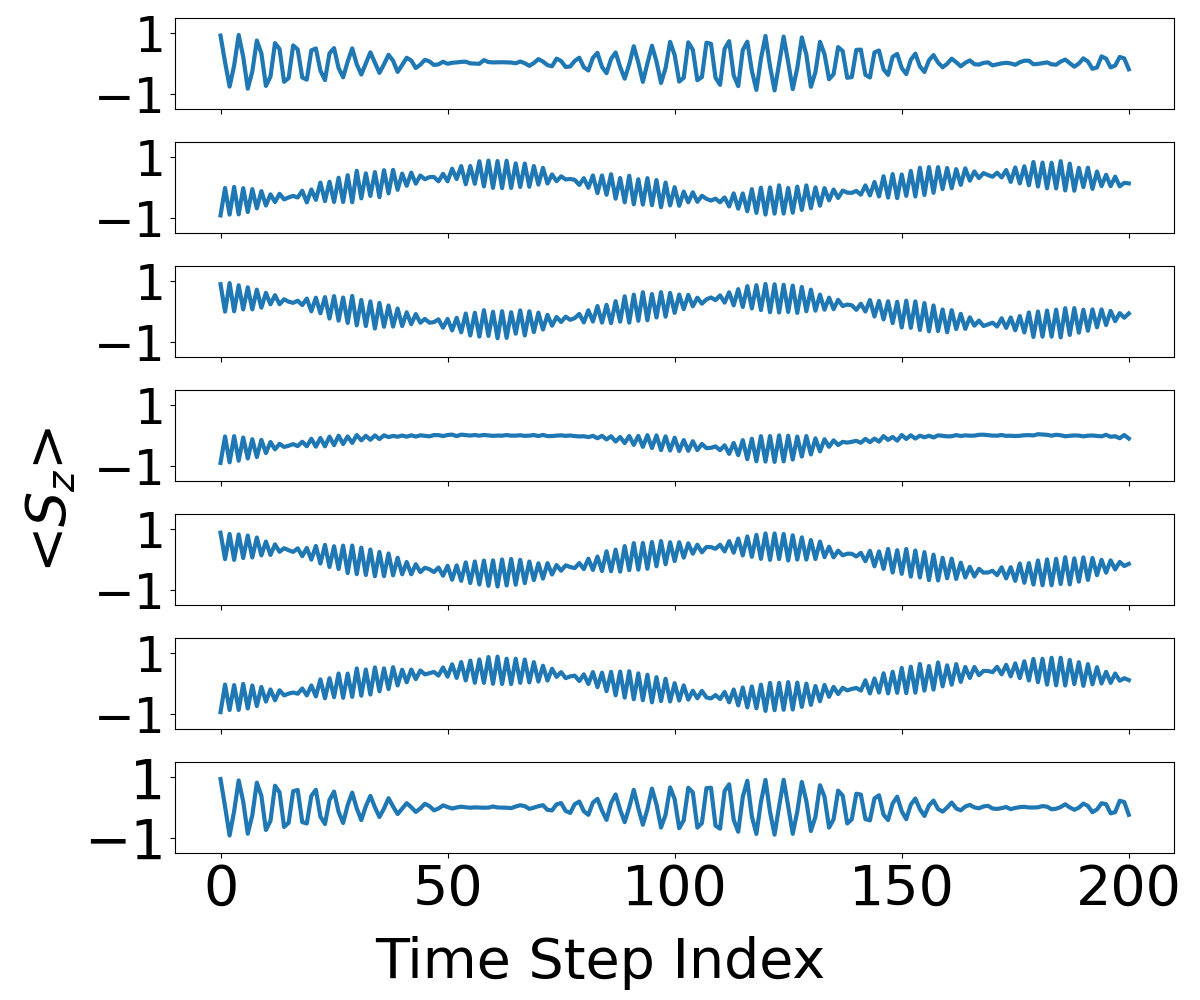}}\hfill
\subcaptionbox{\protect\label{d}}
{\includegraphics[width=0.43\linewidth]{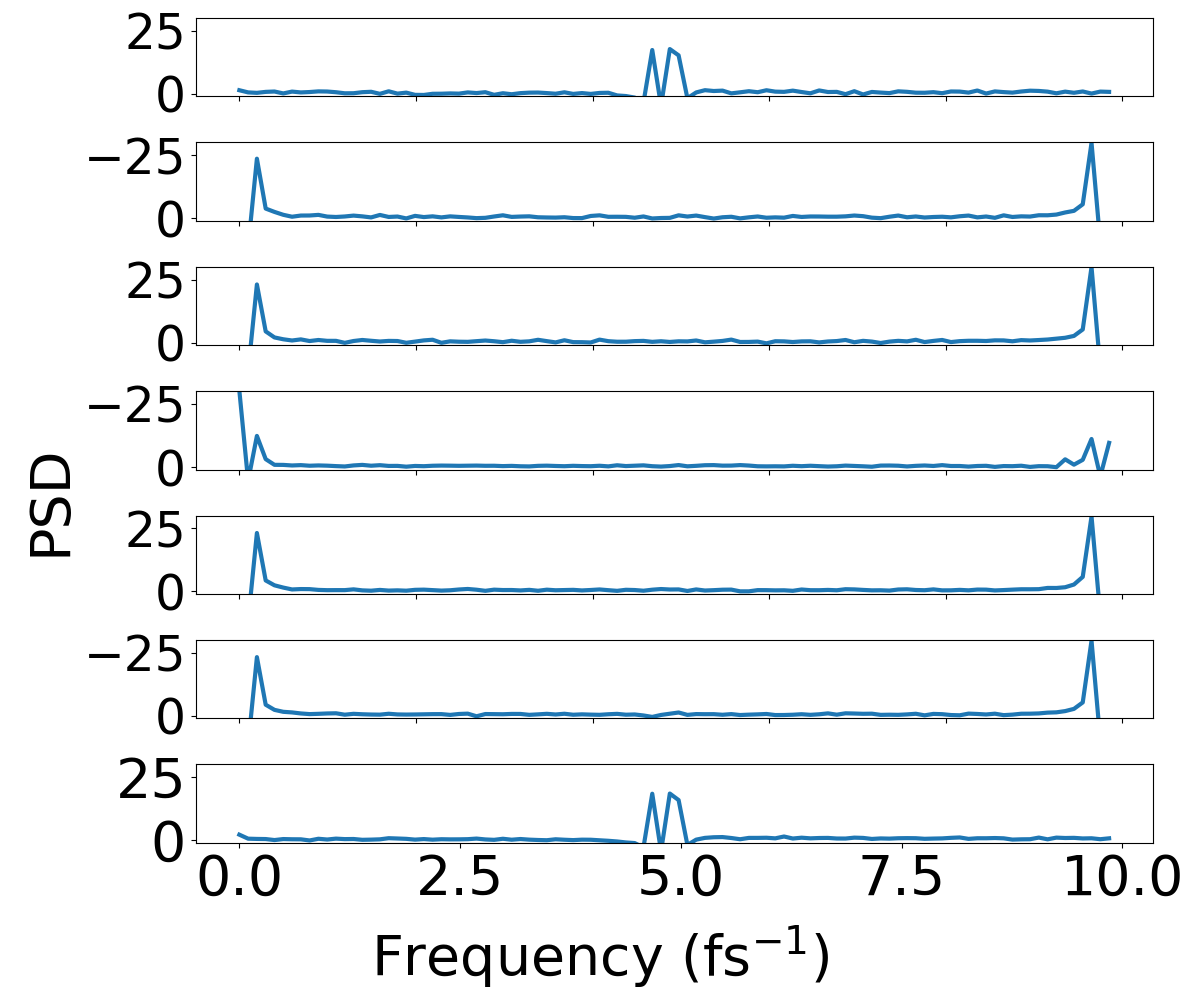}}\hfill
\caption{Time evolution of qubit magnetizations and their corresponding Fourier transforms in power spectral density (PSD). The system is initialized in staggered states along the x- and z-directions respectively.
(a) \( x \)-component of the site-resolved magnetizations for \( J = 0.5\) eV , \( h = 10.0\) eV, and \( \lambda = 1.0 \) (local magnetic field at the central site is changed to  \( h\to (1-\lambda) h \)). Each sub-figure from top to bottom represents individual qubits, from qubit one to qubit seven;  
(b) Fourier transform of the data in (a);  
(c) \( z \)-component of the site-resolved z-magnetizations for \( J = 10.0\) eV, \( h = 0.5 \) eV, and \( \lambda = 1.0 \);  
(d) Fourier transform of the data in (c).  The time step is $\Delta t=0.05 ~\text{fs}$, the total evolution time is $10.0~\text{fs}$.}
\label{figure:results_neel}
\end{figure}

The results are shown in Fig. \ref{figure:results_neel}. In the $h>J$ regime, the system is initialized in a staggered state along the $x$-direction (Figs.~\ref{figure:results_neel}(a),(b)). As the correlation length decreases, the dynamics become increasingly local, with each spin evolving primarily under the transverse field. In contrast, for $J>h$, starting from a staggered state in the $z$-direction, the edge spins exhibit markedly different dynamics than the bulk (Figs.~\ref{figure:results_neel}(c),(d)).This separation of time scales reflects the overlap of boundary spins with the Majorana zero modes, demonstrating that even for quenches from staggered states, clear dynamical signatures of the MZMs persist in the topological phase.

\subsection{Effect of Impurity on the Site-Resolved Time Evolution}

\begin{figure}
\subcaptionbox{\protect\label{a}}{\includegraphics[width=0.43\linewidth]{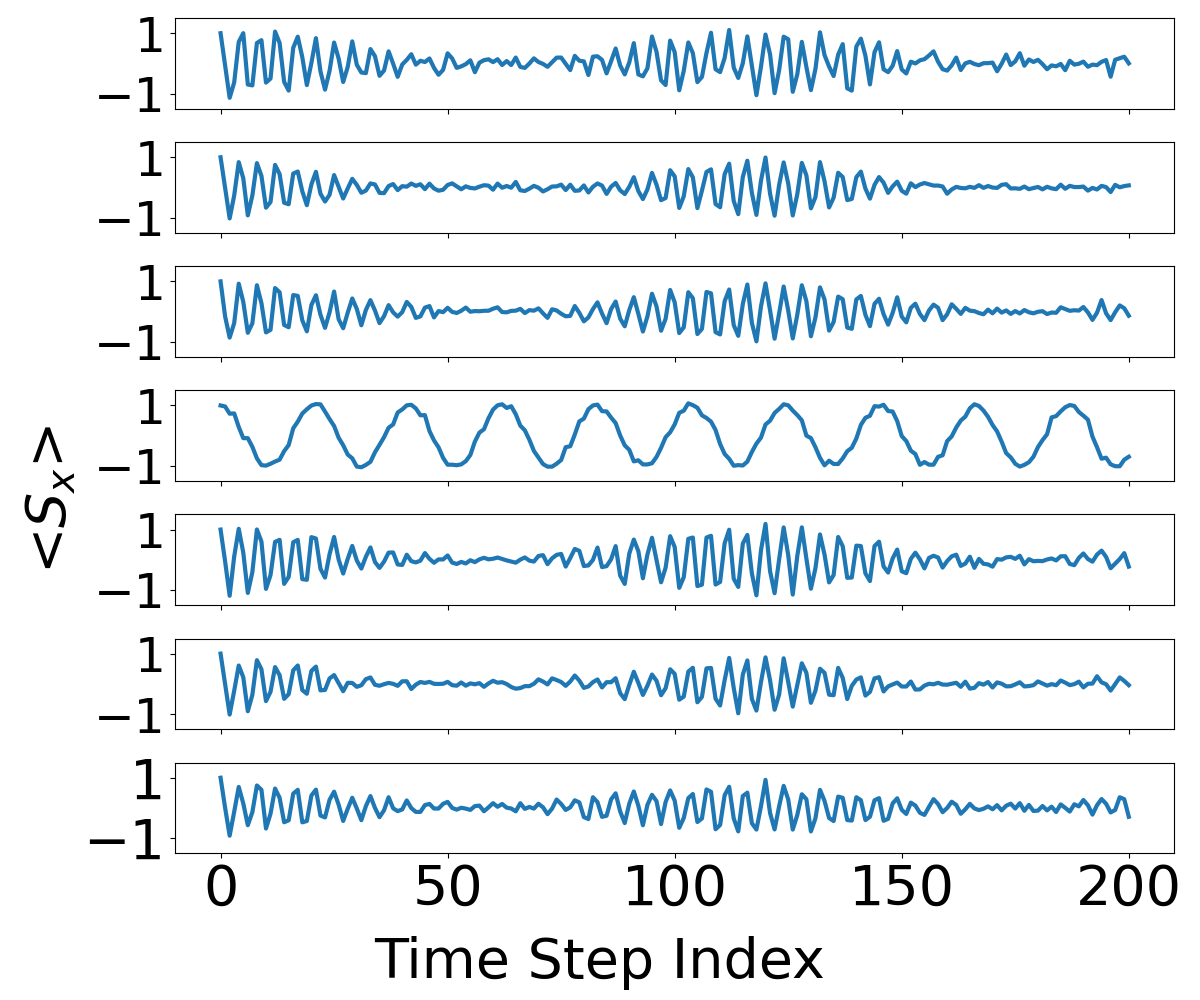}}\hfill
\subcaptionbox{\protect\label{b}}
{\includegraphics[width=0.43\linewidth]{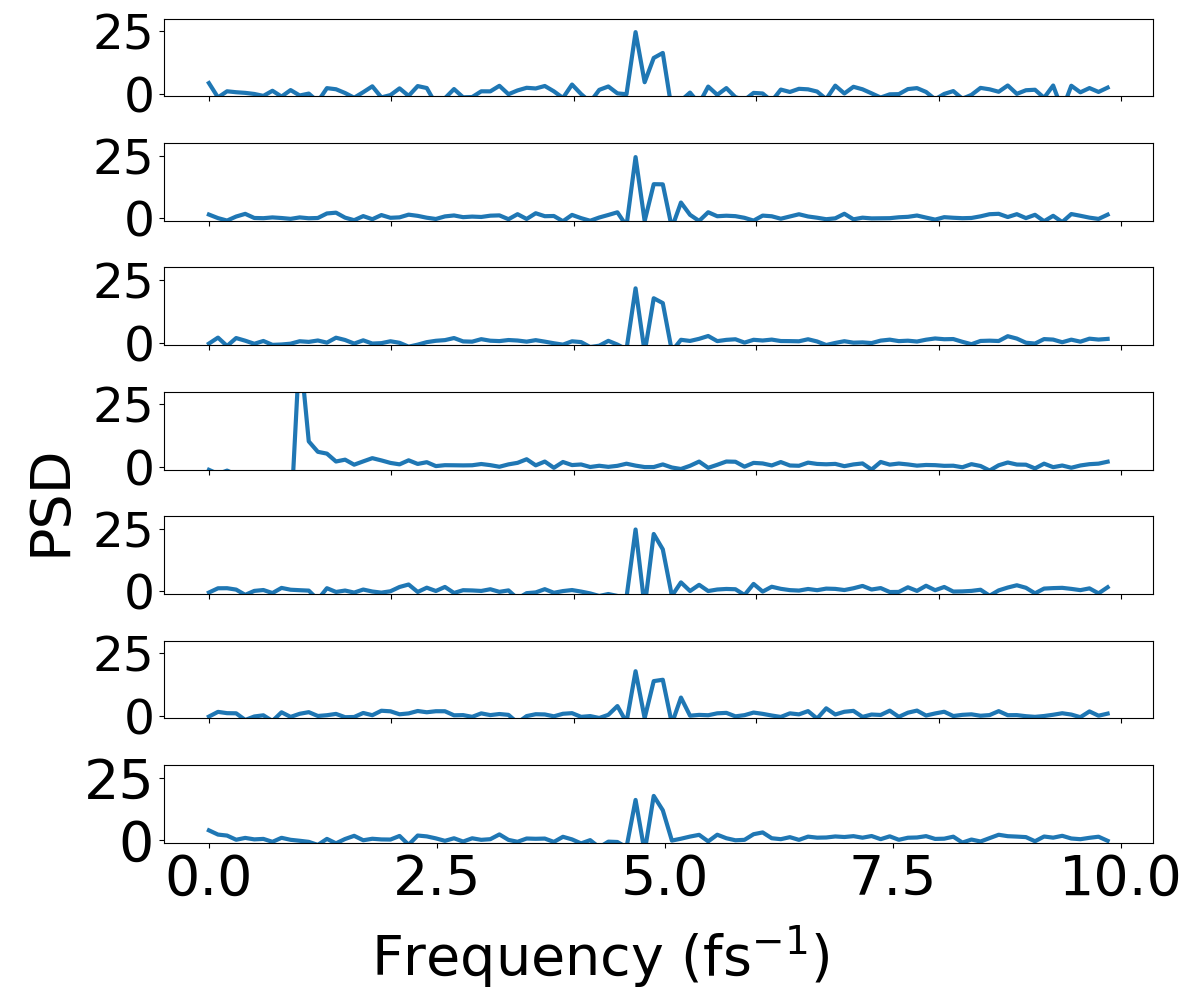}}\hfill
\subcaptionbox{\protect\label{c}}
{\includegraphics[width=0.43\linewidth]{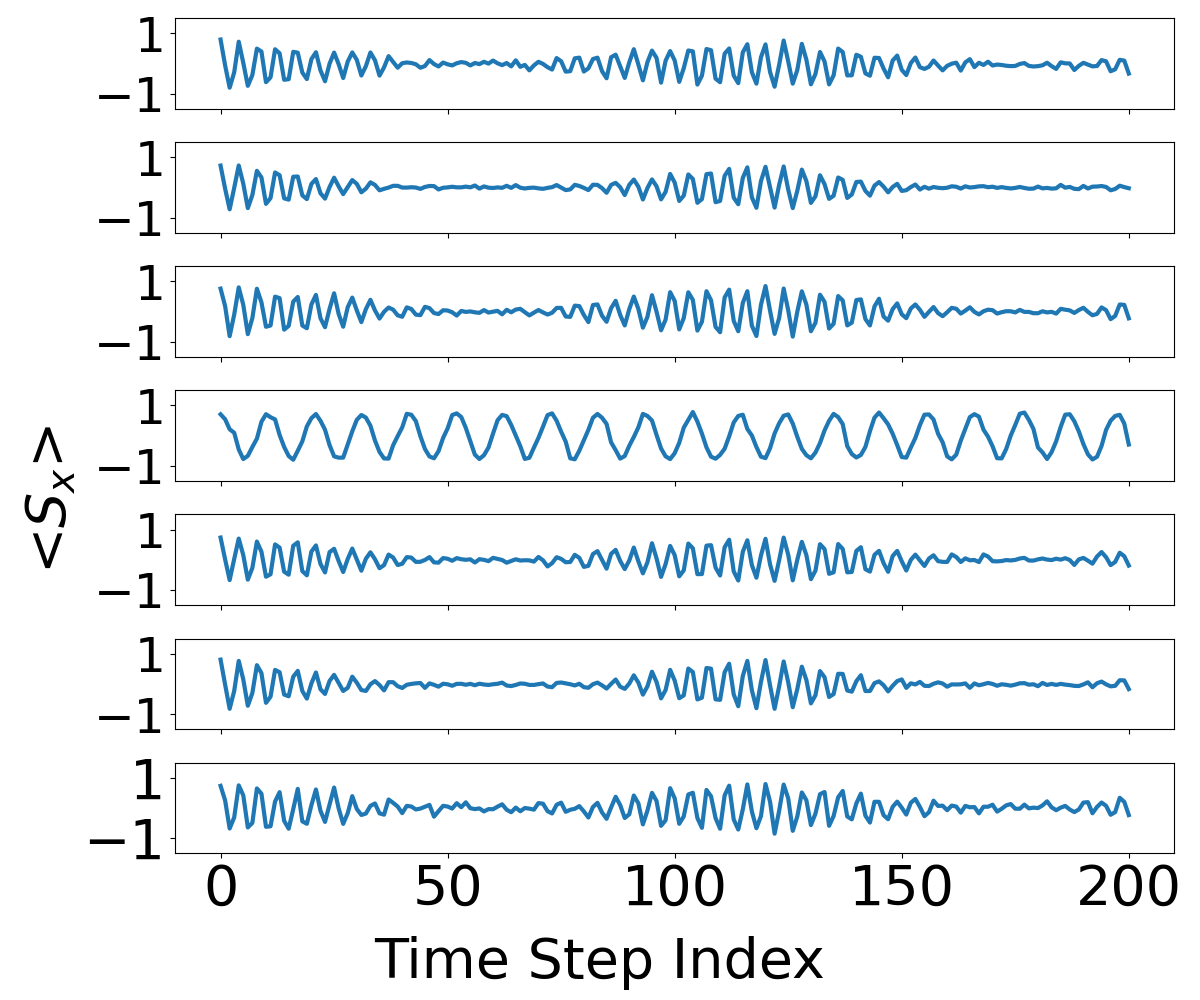}}\hfill
\subcaptionbox{\protect\label{d}}
{\includegraphics[width=0.43\linewidth]{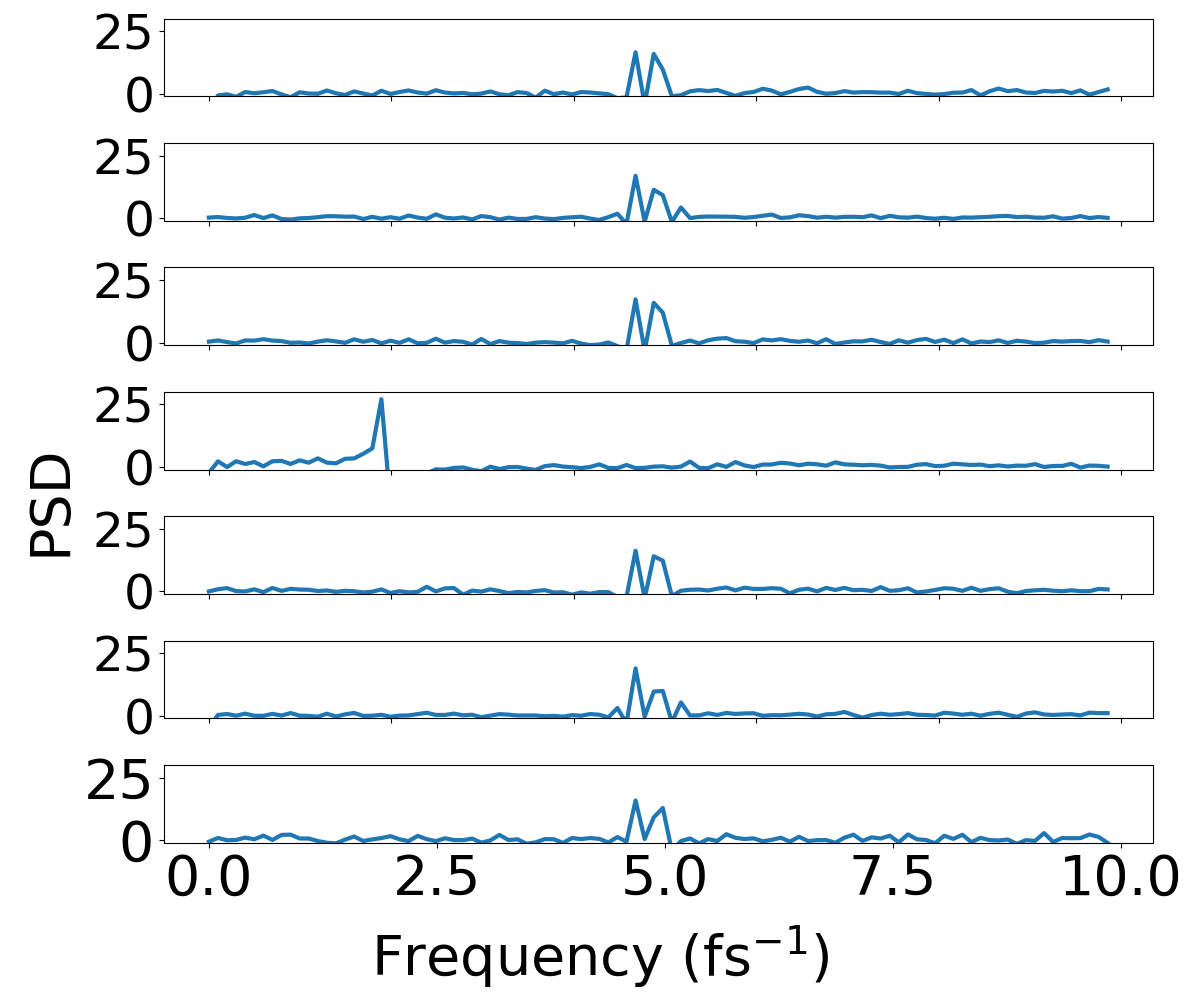}}\hfill
\caption{Time evolution of the \( x \)-component of individual qubit magnetizations in the topologically trivial regime (\( J = 0.5 \) eV, \( h = 10.0\) eV), and an impurity introduced by modifying the local magnetic field at the central site via \( h \to (1-\lambda) h \), with different impurity strengths, and $\Delta t=0.05~\text{fs}$. The system is initialized in the fully polarized state along the x-axis.
(a) \( \lambda = 0.80 \);  
(b) Fourier transform of (a);  
(c) \( \lambda = 0.60 \);  
(d) Fourier transform of (c); 
}
\label{figure:impurity_effect_trivial}
\end{figure}

In the topologically trivial regime (\( h > J \)), the site dynamics are dominated by the local magnetic field. A local impurity therefore primarily alters the behavior of its host site, leading to a distinct evolution relative to the rest of the chain, as shown in Fig. \ref{figure:impurity_effect_trivial}(a-d) for two different impurity strengths.

As the impurity strength is decreased by 25\%, from $\lambda=0.8$ to $\lambda=0.6$, the peak oscillation frequency at the defect site increases from 0.995 to 1.89, indicating greater dynamical recruitment at weaker impurity strength. Compared to the case of maximal impurity strength ($\lambda=1.0$ in Figs. \ref{figure:results_extreme_parameters}(a),(b)), the isolated dynamics of the impurity site gradually blend with those of the remainder of the chain. This trend is clearly visible in both the time series and corresponding Fourier signals: 
%
the dominant frequency at the impurity site shifts toward that of a representative non-impurity site (site 1) as the impurity strength decreases. This convergence is quantified in Fig. \ref{figure:heatmap}(b), which shows the $\delta{\tilde{\omega}}$  frequency difference as a function of the impurity strength $\lambda$, revealing a 
linear trend.

\subsection{Impurity as a Barrier to Excitation Propagation}

 As illustrated in Fig. \ref{figure:model}(b), the parameter \( h \) can be interpreted as the intra-site coupling between two Majorana fermions. In general, an impurity in a spin chain can act as a potential barrier \cite{Kattel_PRB2025,Holger_JPCM1997,Vovrosh_PRB2022}, with the scattering and reflection of excitations depending on the strength of the impurity. Based on this analogy, we propose that a local impurity in \( h \) can serve as a barrier that inhibits the propagation of excitations.

To demonstrate this, we considered a model with \( J = 1.0\) eV and \( h = 4.0\) eV, and initialized the system in the state \( \ket{01111111} \), where the first (edge) spin points along the \( -z \) direction and all other spins point along \( +z \). Under the influence of the coupling \( J \), the local excitation at the edge is expected to propagate into the bulk and affect the magnetization of nearby sites,  behaving like a wave.

\begin{figure}
\subcaptionbox{\protect\label{a}}{\includegraphics[width=0.48\linewidth,height=0.42\linewidth]{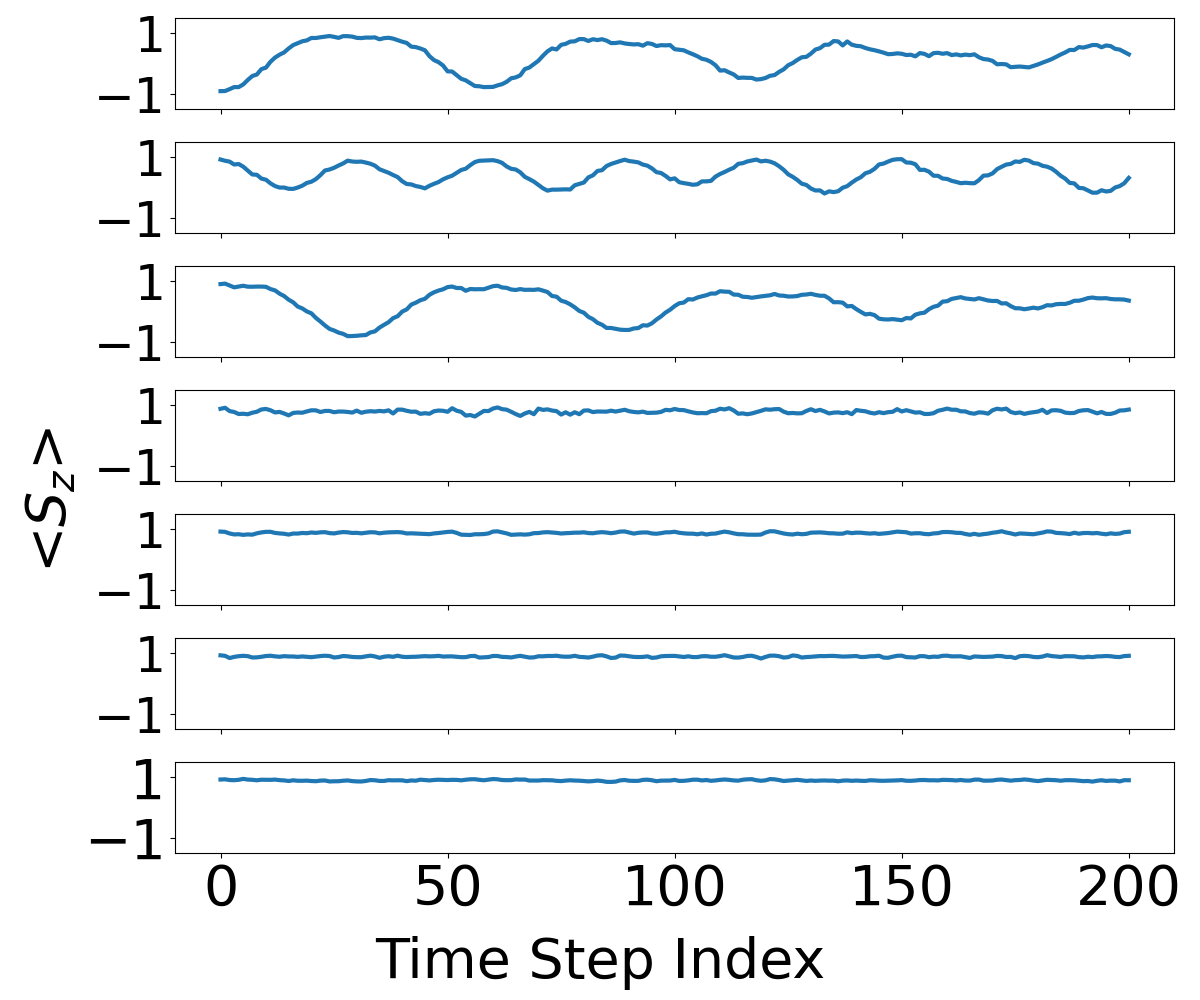}}\hfill
\subcaptionbox{\protect\label{b}}
{\includegraphics[width=0.48\linewidth,height=0.42\linewidth]{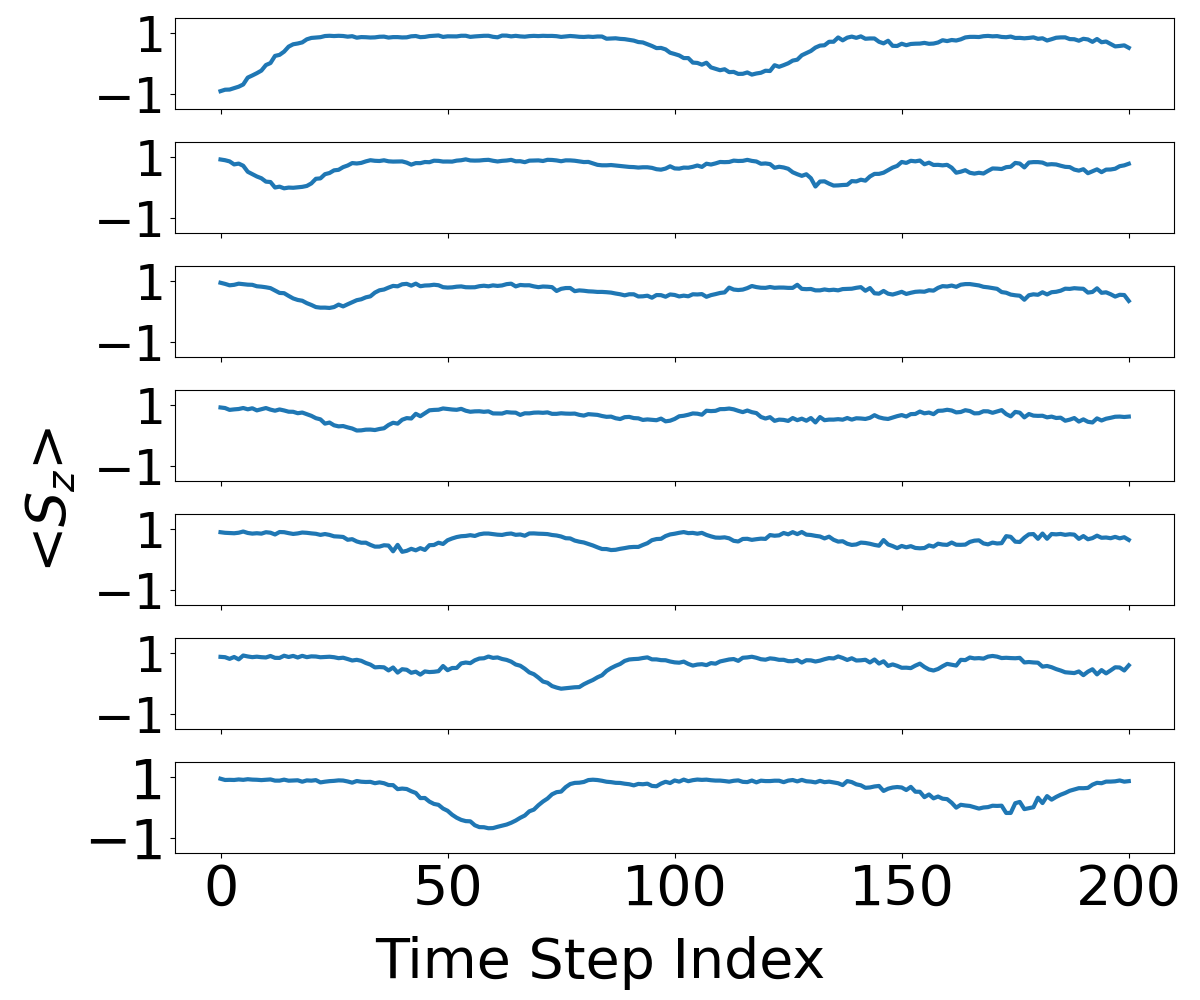}}\hfill
\centering
\subcaptionbox{\protect\label{c}}
{\includegraphics[width=0.48\linewidth,height=0.42\linewidth]{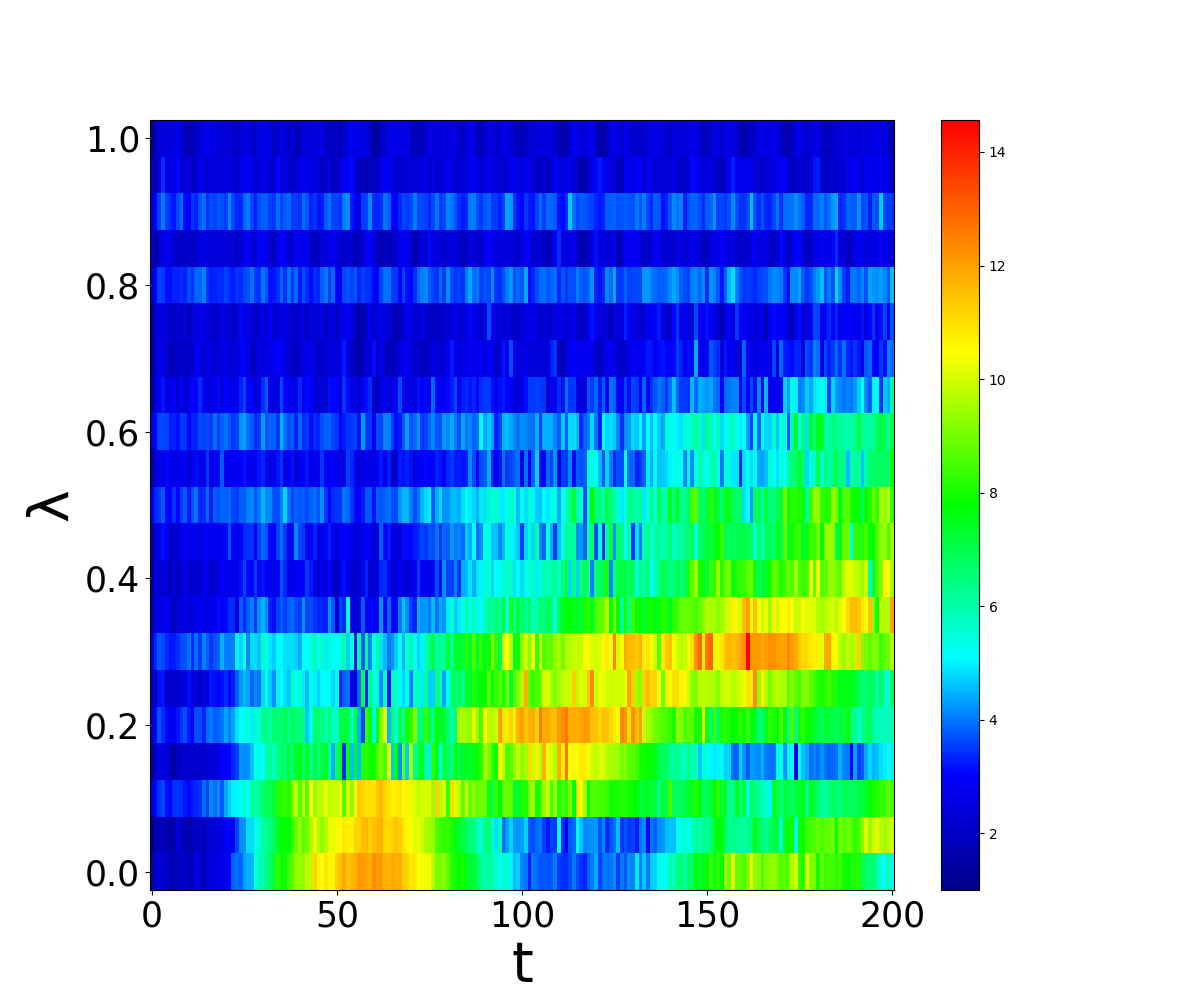}}\hfill
\caption{Time evolution of the site-resolved magnetizations for a 7-qubit chain initialized in the state \( \ket{0111111} \), with coupling \( J = 1.0 \) eV and magnetic field \( h = 4.0\) eV, and an impurity introduced by modifying the local magnetic field at the central site via \( h \to (1-\lambda) h \), shown for different impurity strengths:  
(a) \( \lambda = 1.0 \),  
(b) \( \lambda = 0.0 \).
(c) Heat map of excitation displacement as a function of impurity potential $\lambda$ and time t
The time evolution is obtained by discretizing time with  steps $\Delta t=0.05~\text{fs}$. }  
\label{figure:excitation_impurity_barrier}
\end{figure}

The presence of an impurity in the center of the chain acts as a barrier to this propagation. Intuitively, the stronger the impurity (parameterized by \( \lambda \), Eq. \ref{eq:tfim_hamiltonian}), the more difficult it becomes for the excitation to penetrate through it. The numerical results shown in Fig. \ref{figure:excitation_impurity_barrier} support this picture. For $\lambda=1.0$ the edge excitation at site 1 propagates to sites 2 and 3, and then it is abruptly quenched at site 4 and beyond. As we decrease the impurity strength from $\lambda=1.0$ (Fig. \ref{figure:excitation_impurity_barrier}(a)) to $\lambda=0.7,~0.4,~0.0$ (Figs. \ref{figure:excitation_impurity_barrier}(b),(c),(d)), the sites beyond the impurity are increasingly affected by the wave, indicating partial/complete transmission. 

To further quantify how much the excitation has propagated into the chain beyond the impurity site, the excitation displacement\cite{bassman2022arqtic,K_kc__2022} beyond the impurity site, is measured,
\begin{equation}
    \sigma_{disp} = \sum_{i=\frac{N}{2}}^{N}(i-1)(1-\sigma^z).
    \label{excit_disp}
\end{equation}
A heat map of $\sigma_{disp}$ for different $\lambda$ and times is shown in Fig.  \ref{figure:excitation_impurity_barrier}(c). From the heat map, it is clear that as $\lambda$ goes down, the excitation propagates farther into the chain beyond the impurity site at earlier times, and the excitation enters and exits the region with a higher frequency, which  demonstrates the localization effect of the impurity.

\subsection{Detecting Majorana Fermions in the Center of the Chain}

Thus far, we have  focused on Majorana fermions at the edges of the chain and how they influence the time evolution of edge qubits. To further explore the properties of Majorana fermions, we now turn to the center of the chain and examine how localized Majorana modes affect the evolution of qubits in that region.

Instead of placing an impurity in the on-site magnetic field \( h \), we introduce an impurity in the coupling \( J \) between the fourth and fifth lattice sites. The modified Hamiltonian is then given by

\begin{equation}
    H = J \sum_{\substack{n=1 \\ n \ne N/2+1}}^{N} \sigma^x_n \sigma^x_{n+1} 
    + h \sum_{n=1}^{N+1} \sigma^z_n 
    + J(1 - \lambda) \, \sigma^x_{N/2+1} \sigma^x_{N/2 + 2},
\end{equation}

where the impurity strength is controlled by the parameter \( \lambda \), and the schematic layout is illustrated in Fig. \ref{figure:model} (c). The system is initialized in the state \( \ket{1111111} \).

In the simulations, we set \( J = 10.0\) eV, \( h = 0.5 \) ~eV, and \( \lambda = 1.0 \). This effectively severs the chain at the midpoint, splitting it into two disconnected segments. As a result, two Majorana fermions emerge at the cut: one pair localized at the left edge and at site 4, and another pair localized at site 5 and the right edge.

\begin{figure}
\subcaptionbox{\protect\label{a}}{\includegraphics[width=0.48\linewidth,height=0.42\linewidth]{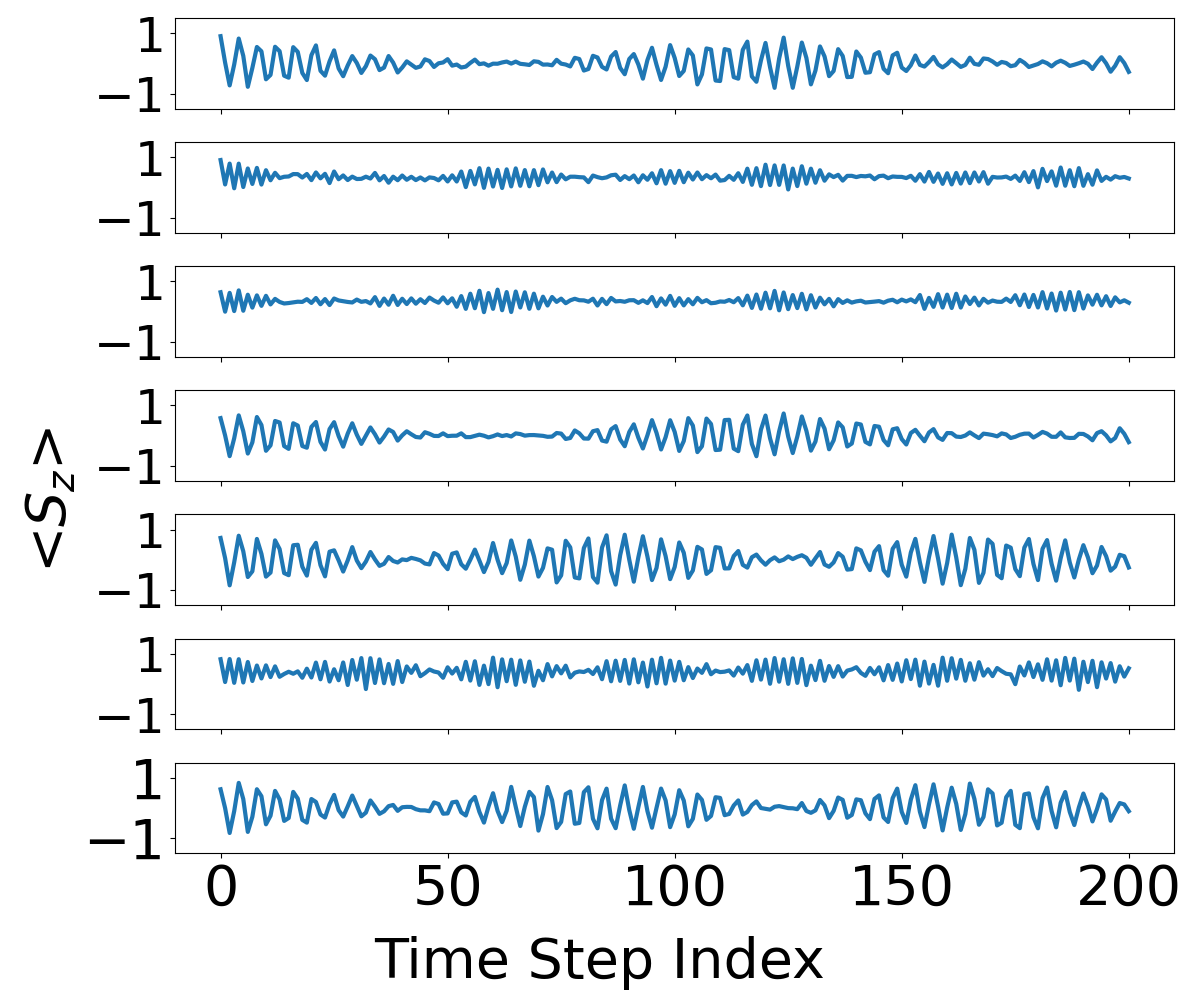}}\hfill
\subcaptionbox{\protect\label{b}}
{\includegraphics[width=0.48\linewidth,height=0.42\linewidth]{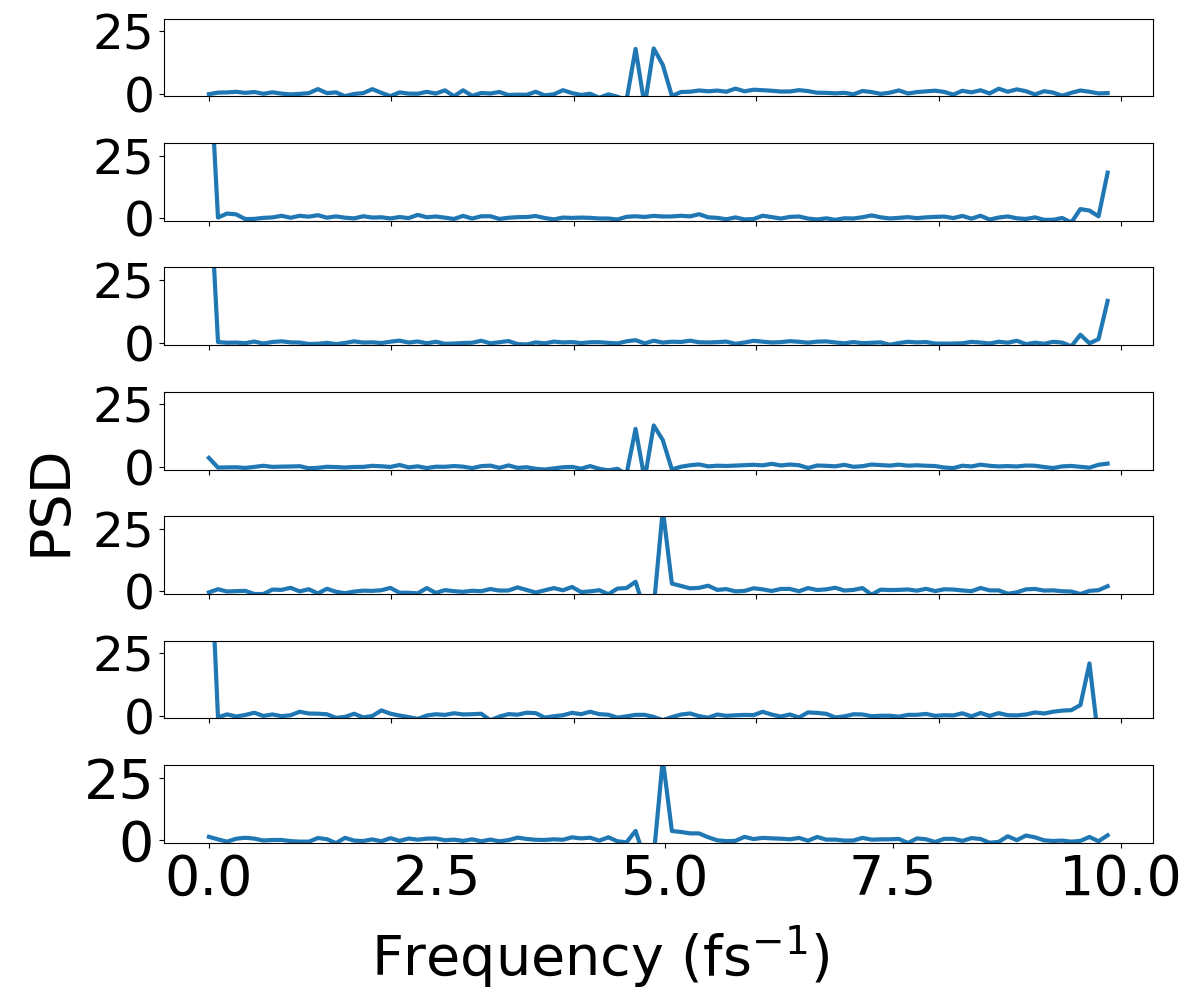}}\hfill
\caption{Time evolution of local magnetizations for a 7-qubit chain with strong coupling, \( J = 10.0\) eV, weak transverse field, \( h = 0.5 \) eV, and an impurity introduced by modifying the bond between sites 4 and 5 via \( J \to (1-\lambda) J \), with \( \lambda = 1.0 \). The system is initialized in the fully polarized state along the z-axis, \( \ket{1111111} \).  
(a) Time evolution of the magnetization at each site.  
(b) Fourier spectra of the site-resolved magnetization dynamics.}
\label{figure:majorana_center}
\end{figure}

From this configuration, we expect the dynamics of sites 1 and 4 to resemble each other, as they are linked by an edge mode within the left sub-chain. Similarly, sites 5 and 7 should exhibit comparable behavior within the right sub-chain. The numerical results confirm this expectation in Fig.~\ref{figure:majorana_center}, showing similar time evolution patterns for site pairs (1, 4) and (5, 7), consistent with the presence of localized Majorana fermions at the internal boundary.

\section{Conclusions}

In this work, we used the Constant Depth Circuit algorithm to investigate the real-time dynamics of a transverse-field Ising chain on noisy intermediate-scale quantum hardware, specifically \texttt{ibm\_kyiv}. 
By representing each spin site as a pair of Majorana fermions with intra-cell coupling \( h \) and inter-cell coupling \( J \), our experiments identify two distinct dynamical regimes governed by the relative strength of \( J \) and \( h \). 
In the regime \( h > J \), the dynamics are dominated by local magnetic fields, and the evolution of each spin site is governed by intra-cell coupling. Conversely, in the regime \( J > h \), the inter-cell coupling is dominant, leading to the effective decoupling of edge Majorana fermions from the bulk and resulting in qualitatively different evolution between edge and bulk sites. By analyzing quenches from both fully polarized and staggered initial states, we reveal the dynamical signatures of the Majorana zero modes and demonstrate their robustness, even when the system is quenched from highly excited states. We also found that a central on-site impurity quenches the edge Majorana modes in the weak-coupling regime, and
the impurity strength tunes the frequency difference between bulk and impurity modes. Furthermore, the central impurity acts as a confinement knob. 

To probe the role of mid-chain Majorana modes, we also introduced an impurity in the coupling \( J \) immediately to the right of the center of the chain. In the \( J < h \) regime, the impurity has little effect, consistent with the delocalized nature of bulk excitations. 
In contrast, in the \( h < J \) regime, the impurity significantly alters the dynamics at and near its location. By initiating spin excitations at one end of the chain, we demonstrated that the impurity acts as a barrier to propagation.

Our results illustrate that the CDC algorithm is an effective and resource-efficient method for simulating quasi-particle dynamics and capturing signatures of emergent Majorana fermions using present-day quantum processors. Its application is currently limited in size by the number of circuit parameters that need to be optimized, which grow with the number of qubits. This is, to our knowledge, the first demonstration of quasi-particle dynamics on a quantum computer. 

\section{Acknowledgements}
The authors acknowledge the Center for Advanced Research Computing (CARC) at the University of Southern California for providing computing resources that have contributed to the research results reported within this publication. URL: https://carc.usc.edu. \\ 
The authors also acknowledge the use of IBM Quantum services for this work. The views expressed are those of the authors, and do not reflect the official policy or position of IBM or the IBM Quantum team. IBM Quantum resources were accessible to us through the IBM Quantum Innovation Center (QIC) at USC. URL: https://dornsife.usc.edu/qic/. 
\\ 
RDF acknowledges partial support by the Defense Advanced Research Projects Agency (DARPA), Contract No. HR001122C0063, and by the Italian ``Ministero dell'Istruzione, Universit\'a e Ricerca'', PRIN 2022 number 2022W9W423. 
SH acknowledges  support by the Department of Energy (DOE-ASCR) Award DE-SC0026337. AN was supported by NSF grant OAC-2118061.


\bibliography{apssamp,Rosa}

\begin{thebibliography}{45}%
\makeatletter
\providecommand \@ifxundefined [1]{%
 \@ifx{#1\undefined}
}%
\providecommand \@ifnum [1]{%
 \ifnum #1\expandafter \@firstoftwo
 \else \expandafter \@secondoftwo
 \fi
}%
\providecommand \@ifx [1]{%
 \ifx #1\expandafter \@firstoftwo
 \else \expandafter \@secondoftwo
 \fi
}%
\providecommand \natexlab [1]{#1}%
\providecommand \enquote  [1]{``#1''}%
\providecommand \bibnamefont  [1]{#1}%
\providecommand \bibfnamefont [1]{#1}%
\providecommand \citenamefont [1]{#1}%
\providecommand \href@noop [0]{\@secondoftwo}%
\providecommand \href [0]{\begingroup \@sanitize@url \@href}%
\providecommand \@href[1]{\@@startlink{#1}\@@href}%
\providecommand \@@href[1]{\endgroup#1\@@endlink}%
\providecommand \@sanitize@url [0]{\catcode `\\12\catcode `\$12\catcode `\&12\catcode `\#12\catcode `\^12\catcode `\_12\catcode `\%12\relax}%
\providecommand \@@startlink[1]{}%
\providecommand \@@endlink[0]{}%
\providecommand \url  [0]{\begingroup\@sanitize@url \@url }%
\providecommand \@url [1]{\endgroup\@href {#1}{\urlprefix }}%
\providecommand \urlprefix  [0]{URL }%
\providecommand \Eprint [0]{\href }%
\providecommand \doibase [0]{http://dx.doi.org/}%
\providecommand \selectlanguage [0]{\@gobble}%
\providecommand \bibinfo  [0]{\@secondoftwo}%
\providecommand \bibfield  [0]{\@secondoftwo}%
\providecommand \translation [1]{[#1]}%
\providecommand \BibitemOpen [0]{}%
\providecommand \bibitemStop [0]{}%
\providecommand \bibitemNoStop [0]{.\EOS\space}%
\providecommand \EOS [0]{\spacefactor3000\relax}%
\providecommand \BibitemShut  [1]{\csname bibitem#1\endcsname}%
\let\auto@bib@innerbib\@empty
\bibitem [{\citenamefont {Kane}\ and\ \citenamefont {Mele}(2005)}]{PhysRevLett.95.226801}%
  \BibitemOpen
  \bibfield  {author} {\bibinfo {author} {\bibfnamefont {C.~L.}\ \bibnamefont {Kane}}\ and\ \bibinfo {author} {\bibfnamefont {E.~J.}\ \bibnamefont {Mele}},\ }\bibfield  {title} {\enquote {\bibinfo {title} {Quantum spin hall effect in graphene},}\ }\href {\doibase 10.1103/PhysRevLett.95.226801} {\bibfield  {journal} {\bibinfo  {journal} {Phys. Rev. Lett.}\ }\textbf {\bibinfo {volume} {95}},\ \bibinfo {pages} {226801} (\bibinfo {year} {2005})}\BibitemShut {NoStop}%
\bibitem [{\citenamefont {König}\ \emph {et~al.}(2007)\citenamefont {König}, \citenamefont {Wiedmann}, \citenamefont {Brüne}, \citenamefont {Roth}, \citenamefont {Buhmann}, \citenamefont {Molenkamp}, \citenamefont {Qi},\ and\ \citenamefont {Zhang}}]{doi:10.1126/science.1148047}%
  \BibitemOpen
  \bibfield  {author} {\bibinfo {author} {\bibfnamefont {Markus}\ \bibnamefont {König}}, \bibinfo {author} {\bibfnamefont {Steffen}\ \bibnamefont {Wiedmann}}, \bibinfo {author} {\bibfnamefont {Christoph}\ \bibnamefont {Brüne}}, \bibinfo {author} {\bibfnamefont {Andreas}\ \bibnamefont {Roth}}, \bibinfo {author} {\bibfnamefont {Hartmut}\ \bibnamefont {Buhmann}}, \bibinfo {author} {\bibfnamefont {Laurens~W.}\ \bibnamefont {Molenkamp}}, \bibinfo {author} {\bibfnamefont {Xiao-Liang}\ \bibnamefont {Qi}}, \ and\ \bibinfo {author} {\bibfnamefont {Shou-Cheng}\ \bibnamefont {Zhang}},\ }\bibfield  {title} {\enquote {\bibinfo {title} {Quantum spin hall insulator state in hgte quantum wells},}\ }\href {\doibase 10.1126/science.1148047} {\bibfield  {journal} {\bibinfo  {journal} {Science}\ }\textbf {\bibinfo {volume} {318}},\ \bibinfo {pages} {766--770} (\bibinfo {year} {2007})},\ \Eprint {http://arxiv.org/abs/https://www.science.org/doi/pdf/10.1126/science.1148047}
  {https://www.science.org/doi/pdf/10.1126/science.1148047} \BibitemShut {NoStop}%
\bibitem [{\citenamefont {Hasan}\ and\ \citenamefont {Kane}(2010)}]{RevModPhys.82.3045}%
  \BibitemOpen
  \bibfield  {author} {\bibinfo {author} {\bibfnamefont {M.~Z.}\ \bibnamefont {Hasan}}\ and\ \bibinfo {author} {\bibfnamefont {C.~L.}\ \bibnamefont {Kane}},\ }\bibfield  {title} {\enquote {\bibinfo {title} {Colloquium: Topological insulators},}\ }\href {\doibase 10.1103/RevModPhys.82.3045} {\bibfield  {journal} {\bibinfo  {journal} {Rev. Mod. Phys.}\ }\textbf {\bibinfo {volume} {82}},\ \bibinfo {pages} {3045--3067} (\bibinfo {year} {2010})}\BibitemShut {NoStop}%
\bibitem [{\citenamefont {Qi}\ and\ \citenamefont {Zhang}(2011)}]{RevModPhys.83.1057}%
  \BibitemOpen
  \bibfield  {author} {\bibinfo {author} {\bibfnamefont {Xiao-Liang}\ \bibnamefont {Qi}}\ and\ \bibinfo {author} {\bibfnamefont {Shou-Cheng}\ \bibnamefont {Zhang}},\ }\bibfield  {title} {\enquote {\bibinfo {title} {Topological insulators and superconductors},}\ }\href {\doibase 10.1103/RevModPhys.83.1057} {\bibfield  {journal} {\bibinfo  {journal} {Rev. Mod. Phys.}\ }\textbf {\bibinfo {volume} {83}},\ \bibinfo {pages} {1057--1110} (\bibinfo {year} {2011})}\BibitemShut {NoStop}%
\bibitem [{\citenamefont {Elliott}\ and\ \citenamefont {Franz}(2015)}]{Elliott_RMP2015_majorana}%
  \BibitemOpen
  \bibfield  {author} {\bibinfo {author} {\bibfnamefont {Steven~R.}\ \bibnamefont {Elliott}}\ and\ \bibinfo {author} {\bibfnamefont {Marcel}\ \bibnamefont {Franz}},\ }\bibfield  {title} {\enquote {\bibinfo {title} {Colloquium: Majorana fermions in nuclear, particle, and solid-state physics},}\ }\href {\doibase 10.1103/RevModPhys.87.137} {\bibfield  {journal} {\bibinfo  {journal} {Rev. Mod. Phys.}\ }\textbf {\bibinfo {volume} {87}},\ \bibinfo {pages} {137--163} (\bibinfo {year} {2015})}\BibitemShut {NoStop}%
\bibitem [{\citenamefont {Kitaev}(2001)}]{Kitaev_2001}%
  \BibitemOpen
  \bibfield  {author} {\bibinfo {author} {\bibfnamefont {A~Yu}\ \bibnamefont {Kitaev}},\ }\bibfield  {title} {\enquote {\bibinfo {title} {Unpaired majorana fermions in quantum wires},}\ }\href {\doibase 10.1070/1063-7869/44/10s/s29} {\bibfield  {journal} {\bibinfo  {journal} {Physics-Uspekhi}\ }\textbf {\bibinfo {volume} {44}},\ \bibinfo {pages} {131–136} (\bibinfo {year} {2001})}\BibitemShut {NoStop}%
\bibitem [{\citenamefont {Romito}\ \emph {et~al.}(2012)\citenamefont {Romito}, \citenamefont {Alicea}, \citenamefont {Refael},\ and\ \citenamefont {von Oppen}}]{PhysRevB.85.020502}%
  \BibitemOpen
  \bibfield  {author} {\bibinfo {author} {\bibfnamefont {Alessandro}\ \bibnamefont {Romito}}, \bibinfo {author} {\bibfnamefont {Jason}\ \bibnamefont {Alicea}}, \bibinfo {author} {\bibfnamefont {Gil}\ \bibnamefont {Refael}}, \ and\ \bibinfo {author} {\bibfnamefont {Felix}\ \bibnamefont {von Oppen}},\ }\bibfield  {title} {\enquote {\bibinfo {title} {Manipulating majorana fermions using supercurrents},}\ }\href {\doibase 10.1103/PhysRevB.85.020502} {\bibfield  {journal} {\bibinfo  {journal} {Phys. Rev. B}\ }\textbf {\bibinfo {volume} {85}},\ \bibinfo {pages} {020502} (\bibinfo {year} {2012})}\BibitemShut {NoStop}%
\bibitem [{\citenamefont {Calabrese}\ and\ \citenamefont {Cardy}(2004)}]{Pasquale_Calabrese_2004}%
  \BibitemOpen
  \bibfield  {author} {\bibinfo {author} {\bibfnamefont {Pasquale}\ \bibnamefont {Calabrese}}\ and\ \bibinfo {author} {\bibfnamefont {John}\ \bibnamefont {Cardy}},\ }\bibfield  {title} {\enquote {\bibinfo {title} {Entanglement entropy and quantum field theory},}\ }\href {\doibase 10.1088/1742-5468/2004/06/P06002} {\bibfield  {journal} {\bibinfo  {journal} {Journal of Statistical Mechanics: Theory and Experiment}\ }\textbf {\bibinfo {volume} {2004}},\ \bibinfo {pages} {P06002} (\bibinfo {year} {2004})}\BibitemShut {NoStop}%
\bibitem [{\citenamefont {Calabrese}\ and\ \citenamefont {Cardy}(2005)}]{Calabrese_2005}%
  \BibitemOpen
  \bibfield  {author} {\bibinfo {author} {\bibfnamefont {Pasquale}\ \bibnamefont {Calabrese}}\ and\ \bibinfo {author} {\bibfnamefont {John}\ \bibnamefont {Cardy}},\ }\bibfield  {title} {\enquote {\bibinfo {title} {Evolution of entanglement entropy in one-dimensional systems},}\ }\href {\doibase 10.1088/1742-5468/2005/04/P04010} {\bibfield  {journal} {\bibinfo  {journal} {Journal of Statistical Mechanics: Theory and Experiment}\ }\textbf {\bibinfo {volume} {2005}},\ \bibinfo {pages} {P04010} (\bibinfo {year} {2005})}\BibitemShut {NoStop}%
\bibitem [{\citenamefont {Calabrese}(2020)}]{10.21468/SciPostPhysLectNotes.20}%
  \BibitemOpen
  \bibfield  {author} {\bibinfo {author} {\bibfnamefont {Pasquale}\ \bibnamefont {Calabrese}},\ }\bibfield  {title} {\enquote {\bibinfo {title} {{Entanglement spreading in non-equilibrium integrable systems}},}\ }\href {\doibase 10.21468/SciPostPhysLectNotes.20} {\bibfield  {journal} {\bibinfo  {journal} {SciPost Phys. Lect. Notes}\ ,\ \bibinfo {pages} {20}} (\bibinfo {year} {2020})}\BibitemShut {NoStop}%
\bibitem [{\citenamefont {Bastianello}\ and\ \citenamefont {Calabrese}(2018)}]{10.21468/SciPostPhys.5.4.033}%
  \BibitemOpen
  \bibfield  {author} {\bibinfo {author} {\bibfnamefont {Alvise}\ \bibnamefont {Bastianello}}\ and\ \bibinfo {author} {\bibfnamefont {Pasquale}\ \bibnamefont {Calabrese}},\ }\bibfield  {title} {\enquote {\bibinfo {title} {{Spreading of entanglement and correlations after a quench with intertwined quasiparticles}},}\ }\href {\doibase 10.21468/SciPostPhys.5.4.033} {\bibfield  {journal} {\bibinfo  {journal} {SciPost Phys.}\ }\textbf {\bibinfo {volume} {5}},\ \bibinfo {pages} {033} (\bibinfo {year} {2018})}\BibitemShut {NoStop}%
\bibitem [{\citenamefont {Sengupta}\ \emph {et~al.}(2004)\citenamefont {Sengupta}, \citenamefont {Powell},\ and\ \citenamefont {Sachdev}}]{PhysRevA.69.053616}%
  \BibitemOpen
  \bibfield  {author} {\bibinfo {author} {\bibfnamefont {K.}~\bibnamefont {Sengupta}}, \bibinfo {author} {\bibfnamefont {Stephen}\ \bibnamefont {Powell}}, \ and\ \bibinfo {author} {\bibfnamefont {Subir}\ \bibnamefont {Sachdev}},\ }\bibfield  {title} {\enquote {\bibinfo {title} {Quench dynamics across quantum critical points},}\ }\href {\doibase 10.1103/PhysRevA.69.053616} {\bibfield  {journal} {\bibinfo  {journal} {Phys. Rev. A}\ }\textbf {\bibinfo {volume} {69}},\ \bibinfo {pages} {053616} (\bibinfo {year} {2004})}\BibitemShut {NoStop}%
\bibitem [{\citenamefont {Mercado}\ \emph {et~al.}(2024{\natexlab{a}})\citenamefont {Mercado}, \citenamefont {Chen}, \citenamefont {Darekar}, \citenamefont {Nakano}, \citenamefont {Di~Felice},\ and\ \citenamefont {Haas}}]{PhysRevB.110.075116}%
  \BibitemOpen
  \bibfield  {author} {\bibinfo {author} {\bibfnamefont {Miguel}\ \bibnamefont {Mercado}}, \bibinfo {author} {\bibfnamefont {Kyle}\ \bibnamefont {Chen}}, \bibinfo {author} {\bibfnamefont {Parth~Hemant}\ \bibnamefont {Darekar}}, \bibinfo {author} {\bibfnamefont {Aiichiro}\ \bibnamefont {Nakano}}, \bibinfo {author} {\bibfnamefont {Rosa}\ \bibnamefont {Di~Felice}}, \ and\ \bibinfo {author} {\bibfnamefont {Stephan}\ \bibnamefont {Haas}},\ }\bibfield  {title} {\enquote {\bibinfo {title} {Dynamics of symmetry-protected topological matter on a quantum computer},}\ }\href {\doibase 10.1103/PhysRevB.110.075116} {\bibfield  {journal} {\bibinfo  {journal} {Phys. Rev. B}\ }\textbf {\bibinfo {volume} {110}},\ \bibinfo {pages} {075116} (\bibinfo {year} {2024}{\natexlab{a}})}\BibitemShut {NoStop}%
\bibitem [{\citenamefont {Yu}\ \emph {et~al.}(2023)\citenamefont {Yu}, \citenamefont {Zhao},\ and\ \citenamefont {Wei}}]{PhysRevResearch.5.013183}%
  \BibitemOpen
  \bibfield  {author} {\bibinfo {author} {\bibfnamefont {Hongye}\ \bibnamefont {Yu}}, \bibinfo {author} {\bibfnamefont {Yusheng}\ \bibnamefont {Zhao}}, \ and\ \bibinfo {author} {\bibfnamefont {Tzu-Chieh}\ \bibnamefont {Wei}},\ }\bibfield  {title} {\enquote {\bibinfo {title} {Simulating large-size quantum spin chains on cloud-based superconducting quantum computers},}\ }\href {\doibase 10.1103/PhysRevResearch.5.013183} {\bibfield  {journal} {\bibinfo  {journal} {Phys. Rev. Res.}\ }\textbf {\bibinfo {volume} {5}},\ \bibinfo {pages} {013183} (\bibinfo {year} {2023})}\BibitemShut {NoStop}%
\bibitem [{\citenamefont {Bassman~Oftelie}\ \emph {et~al.}(2022{\natexlab{a}})\citenamefont {Bassman~Oftelie}, \citenamefont {Van~Beeumen}, \citenamefont {Younis}, \citenamefont {Smith}, \citenamefont {Iancu},\ and\ \citenamefont {de~Jong}}]{Bassman_MT2022}%
  \BibitemOpen
  \bibfield  {author} {\bibinfo {author} {\bibfnamefont {L.}~\bibnamefont {Bassman~Oftelie}}, \bibinfo {author} {\bibfnamefont {R.}~\bibnamefont {Van~Beeumen}}, \bibinfo {author} {\bibfnamefont {E.}~\bibnamefont {Younis}}, \bibinfo {author} {\bibfnamefont {E.}~\bibnamefont {Smith}}, \bibinfo {author} {\bibfnamefont {C.}~\bibnamefont {Iancu}}, \ and\ \bibinfo {author} {\bibfnamefont {W.~A.}\ \bibnamefont {de~Jong}},\ }\bibfield  {title} {\enquote {\bibinfo {title} {Constant-depth circuits for dynamic simulations of materials on quantum computers},}\ }\href@noop {} {\bibfield  {journal} {\bibinfo  {journal} {Materials Theory}\ }\textbf {\bibinfo {volume} {6}},\ \bibinfo {pages} {13} (\bibinfo {year} {2022}{\natexlab{a}})}\BibitemShut {NoStop}%
\bibitem [{\citenamefont {Lancaster}\ and\ \citenamefont {Allen}(2025)}]{Lancaster_AJP2025}%
  \BibitemOpen
  \bibfield  {author} {\bibinfo {author} {\bibfnamefont {Jarrett~L.}\ \bibnamefont {Lancaster}}\ and\ \bibinfo {author} {\bibfnamefont {D.~Brysen}\ \bibnamefont {Allen}},\ }\bibfield  {title} {\enquote {\bibinfo {title} {Simulating spin dynamics with quantum computers},}\ }\href@noop {} {\bibfield  {journal} {\bibinfo  {journal} {American Journal of Physics}\ }\textbf {\bibinfo {volume} {93}},\ \bibinfo {pages} {98--109} (\bibinfo {year} {2025})}\BibitemShut {NoStop}%
\bibitem [{\citenamefont {Gandon}\ \emph {et~al.}(2025)\citenamefont {Gandon}, \citenamefont {Baiardi}, \citenamefont {Rossmannek}, \citenamefont {Dobrautz},\ and\ \citenamefont {Tavernelli}}]{Gandon_PRXQ2025}%
  \BibitemOpen
  \bibfield  {author} {\bibinfo {author} {\bibfnamefont {Anthony}\ \bibnamefont {Gandon}}, \bibinfo {author} {\bibfnamefont {Alberto}\ \bibnamefont {Baiardi}}, \bibinfo {author} {\bibfnamefont {Max}\ \bibnamefont {Rossmannek}}, \bibinfo {author} {\bibfnamefont {Werner}\ \bibnamefont {Dobrautz}}, \ and\ \bibinfo {author} {\bibfnamefont {Ivano}\ \bibnamefont {Tavernelli}},\ }\bibfield  {title} {\enquote {\bibinfo {title} {Quantum computing in spin-adapted representations for efficient simulations of spin systems},}\ }\href@noop {} {\bibfield  {journal} {\bibinfo  {journal} {PRX Quantum}\ }\textbf {\bibinfo {volume} {6}},\ \bibinfo {pages} {030306} (\bibinfo {year} {2025})}\BibitemShut {NoStop}%
\bibitem [{\citenamefont {Lötstedt}\ \emph {et~al.}(2024)\citenamefont {Lötstedt}, \citenamefont {Nishi},\ and\ \citenamefont {Yamanouchi}}]{LOTSTEDT202433}%
  \BibitemOpen
  \bibfield  {author} {\bibinfo {author} {\bibfnamefont {Erik}\ \bibnamefont {Lötstedt}}, \bibinfo {author} {\bibfnamefont {Takanori}\ \bibnamefont {Nishi}}, \ and\ \bibinfo {author} {\bibfnamefont {Kaoru}\ \bibnamefont {Yamanouchi}},\ }\bibfield  {title} {\enquote {\bibinfo {title} {Chapter two - simulation of time-dependent quantum dynamics using quantum computers},}\ }in\ \href {\doibase https://doi.org/10.1016/bs.aamop.2024.05.002} {\emph {\bibinfo {booktitle} {Advances in Atomic, Molecular, and Optical Physics}}},\ \bibinfo {series} {Advances In Atomic, Molecular, and Optical Physics}, Vol.~\bibinfo {volume} {73},\ \bibinfo {editor} {edited by\ \bibinfo {editor} {\bibfnamefont {Louis~F.}\ \bibnamefont {DiMauro}}, \bibinfo {editor} {\bibfnamefont {Hélène}\ \bibnamefont {Perrin}}, \ and\ \bibinfo {editor} {\bibfnamefont {Susanne}\ \bibnamefont {Yelin}}}\ (\bibinfo  {publisher} {Academic Press},\ \bibinfo {year} {2024})\ pp.\ \bibinfo {pages} {33--74}\BibitemShut {NoStop}%
\bibitem [{\citenamefont {Monroe}\ \emph {et~al.}(2021)\citenamefont {Monroe}, \citenamefont {Campbell}, \citenamefont {Duan}, \citenamefont {Gong}, \citenamefont {Gorshkov}, \citenamefont {Hess}, \citenamefont {Islam}, \citenamefont {Kim}, \citenamefont {Linke}, \citenamefont {Pagano}, \citenamefont {Richerme}, \citenamefont {Senko},\ and\ \citenamefont {Yao}}]{Monroe_RMP2021}%
  \BibitemOpen
  \bibfield  {author} {\bibinfo {author} {\bibfnamefont {C.}~\bibnamefont {Monroe}}, \bibinfo {author} {\bibfnamefont {W.~C.}\ \bibnamefont {Campbell}}, \bibinfo {author} {\bibfnamefont {L.-M.}\ \bibnamefont {Duan}}, \bibinfo {author} {\bibfnamefont {Z.-X.}\ \bibnamefont {Gong}}, \bibinfo {author} {\bibfnamefont {A.~V.}\ \bibnamefont {Gorshkov}}, \bibinfo {author} {\bibfnamefont {P.~W.}\ \bibnamefont {Hess}}, \bibinfo {author} {\bibfnamefont {R.}~\bibnamefont {Islam}}, \bibinfo {author} {\bibfnamefont {K.}~\bibnamefont {Kim}}, \bibinfo {author} {\bibfnamefont {N.~M.}\ \bibnamefont {Linke}}, \bibinfo {author} {\bibfnamefont {G.}~\bibnamefont {Pagano}}, \bibinfo {author} {\bibfnamefont {P.}~\bibnamefont {Richerme}}, \bibinfo {author} {\bibfnamefont {C.}~\bibnamefont {Senko}}, \ and\ \bibinfo {author} {\bibfnamefont {N.~Y.}\ \bibnamefont {Yao}},\ }\bibfield  {title} {\enquote {\bibinfo {title} {Programmable quantum simulations of spin systems with trapped ions},}\ }\href {\doibase 10.1103/RevModPhys.93.025001}
  {\bibfield  {journal} {\bibinfo  {journal} {Rev. Mod. Phys.}\ }\textbf {\bibinfo {volume} {93}},\ \bibinfo {pages} {025001} (\bibinfo {year} {2021})}\BibitemShut {NoStop}%
\bibitem [{\citenamefont {B.}(2024)}]{Fauseweh_NatureComm2024}%
  \BibitemOpen
  \bibfield  {author} {\bibinfo {author} {\bibfnamefont {Fauseweh}\ \bibnamefont {B.}},\ }\bibfield  {title} {\enquote {\bibinfo {title} {Quantum many-body simulations on digital quantum computers: State-of-the-art and future challenges},}\ }\href@noop {} {\bibfield  {journal} {\bibinfo  {journal} {Nature Communications}\ }\textbf {\bibinfo {volume} {15}},\ \bibinfo {pages} {2123} (\bibinfo {year} {2024})}\BibitemShut {NoStop}%
\bibitem [{\citenamefont {Mercado}\ \emph {et~al.}(2024{\natexlab{b}})\citenamefont {Mercado}, \citenamefont {Chen}, \citenamefont {Darekar}, \citenamefont {Nakano}, \citenamefont {Di~Felice},\ and\ \citenamefont {Haas}}]{Mercado_PRB2024}%
  \BibitemOpen
  \bibfield  {author} {\bibinfo {author} {\bibfnamefont {M.}~\bibnamefont {Mercado}}, \bibinfo {author} {\bibfnamefont {K.}~\bibnamefont {Chen}}, \bibinfo {author} {\bibfnamefont {P.}~\bibnamefont {Darekar}}, \bibinfo {author} {\bibfnamefont {A.}~\bibnamefont {Nakano}}, \bibinfo {author} {\bibfnamefont {R.}~\bibnamefont {Di~Felice}}, \ and\ \bibinfo {author} {\bibfnamefont {S.}~\bibnamefont {Haas}},\ }\bibfield  {title} {\enquote {\bibinfo {title} {Dynamics of symmetry-protected topological matter on a quantum computer},}\ }\href@noop {} {\bibfield  {journal} {\bibinfo  {journal} {Phys. Rev. B}\ }\textbf {\bibinfo {volume} {110}},\ \bibinfo {pages} {075116} (\bibinfo {year} {2024}{\natexlab{b}})}\BibitemShut {NoStop}%
\bibitem [{\citenamefont {Jaiswal}\ \emph {et~al.}(2025)\citenamefont {Jaiswal}, \citenamefont {Lovas},\ and\ \citenamefont {Balents}}]{Balents_PRA2025}%
  \BibitemOpen
  \bibfield  {author} {\bibinfo {author} {\bibfnamefont {Rimika}\ \bibnamefont {Jaiswal}}, \bibinfo {author} {\bibfnamefont {Izabella}\ \bibnamefont {Lovas}}, \ and\ \bibinfo {author} {\bibfnamefont {Leon}\ \bibnamefont {Balents}},\ }\bibfield  {title} {\enquote {\bibinfo {title} {Simulating a quasiparticle on a quantum device},}\ }\href {\doibase 10.1103/PhysRevA.111.012629} {\bibfield  {journal} {\bibinfo  {journal} {Phys. Rev. A}\ }\textbf {\bibinfo {volume} {111}},\ \bibinfo {pages} {012629} (\bibinfo {year} {2025})}\BibitemShut {NoStop}%
\bibitem [{\citenamefont {Narozhny}(2017)}]{Narozhny_SciRep2017}%
  \BibitemOpen
  \bibfield  {author} {\bibinfo {author} {\bibfnamefont {Boris}\ \bibnamefont {Narozhny}},\ }\bibfield  {title} {\enquote {\bibinfo {title} {Majorana fermions in the nonuniform ising-kitaev chain: exact solution},}\ }\href@noop {} {\bibfield  {journal} {\bibinfo  {journal} {Scientific Reports}\ }\textbf {\bibinfo {volume} {7}},\ \bibinfo {pages} {1447} (\bibinfo {year} {2017})}\BibitemShut {NoStop}%
\bibitem [{\citenamefont {Seiberg}\ and\ \citenamefont {Shao}(2024)}]{Seiberg_ScipostPhys2024}%
  \BibitemOpen
  \bibfield  {author} {\bibinfo {author} {\bibfnamefont {Nathan}\ \bibnamefont {Seiberg}}\ and\ \bibinfo {author} {\bibfnamefont {Shu-Heng}\ \bibnamefont {Shao}},\ }\bibfield  {title} {\enquote {\bibinfo {title} {{Majorana chain and Ising model - (non-invertible) translations, anomalies, and emanant symmetries}},}\ }\href {\doibase 10.21468/SciPostPhys.16.3.064} {\bibfield  {journal} {\bibinfo  {journal} {SciPost Phys.}\ }\textbf {\bibinfo {volume} {16}},\ \bibinfo {pages} {064} (\bibinfo {year} {2024})}\BibitemShut {NoStop}%
\bibitem [{\citenamefont {Schmid}\ \emph {et~al.}(2024)\citenamefont {Schmid}, \citenamefont {Penner}, \citenamefont {Yang}, \citenamefont {Glazman},\ and\ \citenamefont {von Oppen}}]{PhysRevLett.132.210401}%
  \BibitemOpen
  \bibfield  {author} {\bibinfo {author} {\bibfnamefont {Harald}\ \bibnamefont {Schmid}}, \bibinfo {author} {\bibfnamefont {Alexander-Georg}\ \bibnamefont {Penner}}, \bibinfo {author} {\bibfnamefont {Kang}\ \bibnamefont {Yang}}, \bibinfo {author} {\bibfnamefont {Leonid}\ \bibnamefont {Glazman}}, \ and\ \bibinfo {author} {\bibfnamefont {Felix}\ \bibnamefont {von Oppen}},\ }\bibfield  {title} {\enquote {\bibinfo {title} {Robust spectral $\ensuremath{\pi}$ pairing in the random-field floquet quantum ising model},}\ }\href {\doibase 10.1103/PhysRevLett.132.210401} {\bibfield  {journal} {\bibinfo  {journal} {Phys. Rev. Lett.}\ }\textbf {\bibinfo {volume} {132}},\ \bibinfo {pages} {210401} (\bibinfo {year} {2024})}\BibitemShut {NoStop}%
\bibitem [{\citenamefont {Schmid}\ \emph {et~al.}(2025)\citenamefont {Schmid}, \citenamefont {Peng}, \citenamefont {Refael},\ and\ \citenamefont {von Oppen}}]{hn66-j8pt}%
  \BibitemOpen
  \bibfield  {author} {\bibinfo {author} {\bibfnamefont {Harald}\ \bibnamefont {Schmid}}, \bibinfo {author} {\bibfnamefont {Yang}\ \bibnamefont {Peng}}, \bibinfo {author} {\bibfnamefont {Gil}\ \bibnamefont {Refael}}, \ and\ \bibinfo {author} {\bibfnamefont {Felix}\ \bibnamefont {von Oppen}},\ }\bibfield  {title} {\enquote {\bibinfo {title} {Self-similar phase diagram of the fibonacci-driven quantum ising model},}\ }\href {\doibase 10.1103/hn66-j8pt} {\bibfield  {journal} {\bibinfo  {journal} {Phys. Rev. Lett.}\ }\textbf {\bibinfo {volume} {134}},\ \bibinfo {pages} {240404} (\bibinfo {year} {2025})}\BibitemShut {NoStop}%
\bibitem [{\citenamefont {Kemp}\ \emph {et~al.}(2017)\citenamefont {Kemp}, \citenamefont {Yao}, \citenamefont {Laumann},\ and\ \citenamefont {Fendley}}]{Kemp_2017}%
  \BibitemOpen
  \bibfield  {author} {\bibinfo {author} {\bibfnamefont {Jack}\ \bibnamefont {Kemp}}, \bibinfo {author} {\bibfnamefont {Norman~Y}\ \bibnamefont {Yao}}, \bibinfo {author} {\bibfnamefont {Christopher~R}\ \bibnamefont {Laumann}}, \ and\ \bibinfo {author} {\bibfnamefont {Paul}\ \bibnamefont {Fendley}},\ }\bibfield  {title} {\enquote {\bibinfo {title} {Long coherence times for edge spins},}\ }\href {\doibase 10.1088/1742-5468/aa73f0} {\bibfield  {journal} {\bibinfo  {journal} {Journal of Statistical Mechanics: Theory and Experiment}\ }\textbf {\bibinfo {volume} {2017}},\ \bibinfo {pages} {063105} (\bibinfo {year} {2017})}\BibitemShut {NoStop}%
\bibitem [{\citenamefont {Fendley}(2012)}]{Fendley_2012}%
  \BibitemOpen
  \bibfield  {author} {\bibinfo {author} {\bibfnamefont {Paul}\ \bibnamefont {Fendley}},\ }\bibfield  {title} {\enquote {\bibinfo {title} {Parafermionic edge zero modes in zn-invariant spin chains},}\ }\href {\doibase 10.1088/1742-5468/2012/11/P11020} {\bibfield  {journal} {\bibinfo  {journal} {Journal of Statistical Mechanics: Theory and Experiment}\ }\textbf {\bibinfo {volume} {2012}},\ \bibinfo {pages} {P11020} (\bibinfo {year} {2012})}\BibitemShut {NoStop}%
\bibitem [{\citenamefont {Preskill}(2018)}]{Preskill2018quantumcomputingin}%
  \BibitemOpen
  \bibfield  {author} {\bibinfo {author} {\bibfnamefont {John}\ \bibnamefont {Preskill}},\ }\bibfield  {title} {\enquote {\bibinfo {title} {Quantum {C}omputing in the {NISQ} era and beyond},}\ }\href@noop {} {\bibfield  {journal} {\bibinfo  {journal} {{Quantum}}\ }\textbf {\bibinfo {volume} {2}},\ \bibinfo {pages} {79} (\bibinfo {year} {2018})}\BibitemShut {NoStop}%
\bibitem [{\citenamefont {Andrei}\ \emph {et~al.}(1983)\citenamefont {Andrei}, \citenamefont {Furuya},\ and\ \citenamefont {Lowenstein}}]{RevModPhys.55.331}%
  \BibitemOpen
  \bibfield  {author} {\bibinfo {author} {\bibfnamefont {N.}~\bibnamefont {Andrei}}, \bibinfo {author} {\bibfnamefont {K.}~\bibnamefont {Furuya}}, \ and\ \bibinfo {author} {\bibfnamefont {J.~H.}\ \bibnamefont {Lowenstein}},\ }\bibfield  {title} {\enquote {\bibinfo {title} {Solution of the kondo problem},}\ }\href {\doibase 10.1103/RevModPhys.55.331} {\bibfield  {journal} {\bibinfo  {journal} {Rev. Mod. Phys.}\ }\textbf {\bibinfo {volume} {55}},\ \bibinfo {pages} {331--402} (\bibinfo {year} {1983})}\BibitemShut {NoStop}%
\bibitem [{\citenamefont {Mourik}\ \emph {et~al.}(2012)\citenamefont {Mourik}, \citenamefont {Zuo}, \citenamefont {Frolov}, \citenamefont {Plissard}, \citenamefont {Bakkers},\ and\ \citenamefont {Kouwenhoven}}]{doi:10.1126/science.1222360}%
  \BibitemOpen
  \bibfield  {author} {\bibinfo {author} {\bibfnamefont {V.}~\bibnamefont {Mourik}}, \bibinfo {author} {\bibfnamefont {K.}~\bibnamefont {Zuo}}, \bibinfo {author} {\bibfnamefont {S.~M.}\ \bibnamefont {Frolov}}, \bibinfo {author} {\bibfnamefont {S.~R.}\ \bibnamefont {Plissard}}, \bibinfo {author} {\bibfnamefont {E.~P. A.~M.}\ \bibnamefont {Bakkers}}, \ and\ \bibinfo {author} {\bibfnamefont {L.~P.}\ \bibnamefont {Kouwenhoven}},\ }\bibfield  {title} {\enquote {\bibinfo {title} {Signatures of majorana fermions in hybrid superconductor-semiconductor nanowire devices},}\ }\href {\doibase 10.1126/science.1222360} {\bibfield  {journal} {\bibinfo  {journal} {Science}\ }\textbf {\bibinfo {volume} {336}},\ \bibinfo {pages} {1003--1007} (\bibinfo {year} {2012})},\ \Eprint {http://arxiv.org/abs/https://www.science.org/doi/pdf/10.1126/science.1222360} {https://www.science.org/doi/pdf/10.1126/science.1222360} \BibitemShut {NoStop}%
\bibitem [{\citenamefont {Tan}\ \emph {et~al.}(2025)\citenamefont {Tan}, \citenamefont {Hang}, \citenamefont {Haas},\ and\ \citenamefont {Saleur}}]{10.21468/SciPostPhysCore.8.1.002}%
  \BibitemOpen
  \bibfield  {author} {\bibinfo {author} {\bibfnamefont {Chunyu}\ \bibnamefont {Tan}}, \bibinfo {author} {\bibfnamefont {Yuxiao}\ \bibnamefont {Hang}}, \bibinfo {author} {\bibfnamefont {Stephan}\ \bibnamefont {Haas}}, \ and\ \bibinfo {author} {\bibfnamefont {Hubert}\ \bibnamefont {Saleur}},\ }\bibfield  {title} {\enquote {\bibinfo {title} {{Entanglement flow in the Kane-Fisher quantum impurity problem}},}\ }\href {\doibase 10.21468/SciPostPhysCore.8.1.002} {\bibfield  {journal} {\bibinfo  {journal} {SciPost Phys. Core}\ }\textbf {\bibinfo {volume} {8}},\ \bibinfo {pages} {002} (\bibinfo {year} {2025})}\BibitemShut {NoStop}%
\bibitem [{\citenamefont {Müller}\ and\ \citenamefont {Nersesyan}(2016)}]{MULLER2016482}%
  \BibitemOpen
  \bibfield  {author} {\bibinfo {author} {\bibfnamefont {Markus}\ \bibnamefont {Müller}}\ and\ \bibinfo {author} {\bibfnamefont {Alexander~A.}\ \bibnamefont {Nersesyan}},\ }\bibfield  {title} {\enquote {\bibinfo {title} {Classical impurities and boundary majorana zero modes in quantum chains},}\ }\href {\doibase https://doi.org/10.1016/j.aop.2016.07.025} {\bibfield  {journal} {\bibinfo  {journal} {Annals of Physics}\ }\textbf {\bibinfo {volume} {372}},\ \bibinfo {pages} {482--522} (\bibinfo {year} {2016})}\BibitemShut {NoStop}%
\bibitem [{\citenamefont {Ul~Haq}\ and\ \citenamefont {Kauffman}(2021)}]{condmat6010011}%
  \BibitemOpen
  \bibfield  {author} {\bibinfo {author} {\bibfnamefont {Rukhsan}\ \bibnamefont {Ul~Haq}}\ and\ \bibinfo {author} {\bibfnamefont {Louis~H.}\ \bibnamefont {Kauffman}},\ }\bibfield  {title} {\enquote {\bibinfo {title} {Z2 topological order and topological protection of majorana fermion qubits},}\ }\href@noop {} {\bibfield  {journal} {\bibinfo  {journal} {Condensed Matter}\ }\textbf {\bibinfo {volume} {6}} (\bibinfo {year} {2021})}\BibitemShut {NoStop}%
\bibitem [{\citenamefont {Suzuki}(1991)}]{TrotterSuzuki1}%
  \BibitemOpen
  \bibfield  {author} {\bibinfo {author} {\bibfnamefont {Masuo}\ \bibnamefont {Suzuki}},\ }\bibfield  {title} {\enquote {\bibinfo {title} {General theory of fractal path integrals with applications to many‐body theories and statistical physics},}\ }\href@noop {} {\bibfield  {journal} {\bibinfo  {journal} {Journal of Mathematical Physics}\ }\textbf {\bibinfo {volume} {32}},\ \bibinfo {pages} {400--407} (\bibinfo {year} {1991})}\BibitemShut {NoStop}%
\bibitem [{\citenamefont {Trotter}(1959{\natexlab{a}})}]{TrotterSuzuki2}%
  \BibitemOpen
  \bibfield  {author} {\bibinfo {author} {\bibfnamefont {H.~F.}\ \bibnamefont {Trotter}},\ }\bibfield  {title} {\enquote {\bibinfo {title} {On the product of semi-groups of operators},}\ }\href@noop {} {\bibfield  {journal} {\bibinfo  {journal} {Proc. Amer. Math. Soc.}\ }\textbf {\bibinfo {volume} {10}},\ \bibinfo {pages} {545--551} (\bibinfo {year} {1959}{\natexlab{a}})}\BibitemShut {NoStop}%
\bibitem [{\citenamefont {Suzuki}(1976)}]{TrotterSuzuki3}%
  \BibitemOpen
  \bibfield  {author} {\bibinfo {author} {\bibfnamefont {Masuo}\ \bibnamefont {Suzuki}},\ }\bibfield  {title} {\enquote {\bibinfo {title} {Generalized trotter's formula and systematic approximants of exponential operators and inner derivations with applications to many-body problems},}\ }\href@noop {} {\bibfield  {journal} {\bibinfo  {journal} {Communications in Mathematical Physics}\ }\textbf {\bibinfo {volume} {51}},\ \bibinfo {pages} {183--190} (\bibinfo {year} {1976})}\BibitemShut {NoStop}%
\bibitem [{\citenamefont {Bassman~Oftelie}\ \emph {et~al.}(2022{\natexlab{b}})\citenamefont {Bassman~Oftelie}, \citenamefont {Van~Beeumen}, \citenamefont {Younis}, \citenamefont {Smith}, \citenamefont {Iancu},\ and\ \citenamefont {de~Jong}}]{Bassman_Oftelie_2022}%
  \BibitemOpen
  \bibfield  {author} {\bibinfo {author} {\bibfnamefont {Lindsay}\ \bibnamefont {Bassman~Oftelie}}, \bibinfo {author} {\bibfnamefont {Roel}\ \bibnamefont {Van~Beeumen}}, \bibinfo {author} {\bibfnamefont {Ed}~\bibnamefont {Younis}}, \bibinfo {author} {\bibfnamefont {Ethan}\ \bibnamefont {Smith}}, \bibinfo {author} {\bibfnamefont {Costin}\ \bibnamefont {Iancu}}, \ and\ \bibinfo {author} {\bibfnamefont {Wibe~A.}\ \bibnamefont {de~Jong}},\ }\bibfield  {title} {\enquote {\bibinfo {title} {Constant-depth circuits for dynamic simulations of materials on quantum computers},}\ }\href {\doibase 10.1186/s41313-022-00043-x} {\bibfield  {journal} {\bibinfo  {journal} {Materials Theory}\ }\textbf {\bibinfo {volume} {6}} (\bibinfo {year} {2022}{\natexlab{b}}),\ 10.1186/s41313-022-00043-x}\BibitemShut {NoStop}%
\bibitem [{\citenamefont {Koopmans}(1995)}]{koopmans1995spectral}%
  \BibitemOpen
  \bibfield  {author} {\bibinfo {author} {\bibfnamefont {Lambert~H}\ \bibnamefont {Koopmans}},\ }\href@noop {} {\emph {\bibinfo {title} {The spectral analysis of time series}}}\ (\bibinfo  {publisher} {Elsevier},\ \bibinfo {year} {1995})\BibitemShut {NoStop}%
\bibitem [{\citenamefont {Kattel}\ \emph {et~al.}(2025)\citenamefont {Kattel}, \citenamefont {Pasnoori}, \citenamefont {Pixley},\ and\ \citenamefont {Andrei}}]{Kattel_PRB2025}%
  \BibitemOpen
  \bibfield  {author} {\bibinfo {author} {\bibfnamefont {Pradip}\ \bibnamefont {Kattel}}, \bibinfo {author} {\bibfnamefont {Parameshwar~R.}\ \bibnamefont {Pasnoori}}, \bibinfo {author} {\bibfnamefont {J.~H.}\ \bibnamefont {Pixley}}, \ and\ \bibinfo {author} {\bibfnamefont {Natan}\ \bibnamefont {Andrei}},\ }\bibfield  {title} {\enquote {\bibinfo {title} {Edge modes and boundary impurities in the anisotropic heisenberg spin chain},}\ }\href {\doibase 10.1103/PhysRevB.111.174430} {\bibfield  {journal} {\bibinfo  {journal} {Phys. Rev. B}\ }\textbf {\bibinfo {volume} {111}},\ \bibinfo {pages} {174430} (\bibinfo {year} {2025})}\BibitemShut {NoStop}%
\bibitem [{\citenamefont {Frahm}\ and\ \citenamefont {Zvyagin}(1997)}]{Holger_JPCM1997}%
  \BibitemOpen
  \bibfield  {author} {\bibinfo {author} {\bibfnamefont {Holger}\ \bibnamefont {Frahm}}\ and\ \bibinfo {author} {\bibfnamefont {Andrei~A}\ \bibnamefont {Zvyagin}},\ }\bibfield  {title} {\enquote {\bibinfo {title} {The open spin chain with impurity: an exact solution},}\ }\href@noop {} {\bibfield  {journal} {\bibinfo  {journal} {Journal of Physics: Condensed Matter}\ }\textbf {\bibinfo {volume} {9}},\ \bibinfo {pages} {9939} (\bibinfo {year} {1997})}\BibitemShut {NoStop}%
\bibitem [{\citenamefont {Vovrosh}\ \emph {et~al.}(2022)\citenamefont {Vovrosh}, \citenamefont {Zhao}, \citenamefont {Knolle},\ and\ \citenamefont {Bastianello}}]{Vovrosh_PRB2022}%
  \BibitemOpen
  \bibfield  {author} {\bibinfo {author} {\bibfnamefont {Joseph}\ \bibnamefont {Vovrosh}}, \bibinfo {author} {\bibfnamefont {Hongzheng}\ \bibnamefont {Zhao}}, \bibinfo {author} {\bibfnamefont {Johannes}\ \bibnamefont {Knolle}}, \ and\ \bibinfo {author} {\bibfnamefont {Alvise}\ \bibnamefont {Bastianello}},\ }\bibfield  {title} {\enquote {\bibinfo {title} {Confinement-induced impurity states in spin chains},}\ }\href {\doibase 10.1103/PhysRevB.105.L100301} {\bibfield  {journal} {\bibinfo  {journal} {Phys. Rev. B}\ }\textbf {\bibinfo {volume} {105}},\ \bibinfo {pages} {L100301} (\bibinfo {year} {2022})}\BibitemShut {NoStop}%
\bibitem [{\citenamefont {Bassman~Oftelie}\ \emph {et~al.}(2022{\natexlab{c}})\citenamefont {Bassman~Oftelie}, \citenamefont {Powers},\ and\ \citenamefont {De~Jong}}]{bassman2022arqtic}%
  \BibitemOpen
  \bibfield  {author} {\bibinfo {author} {\bibfnamefont {Lindsay}\ \bibnamefont {Bassman~Oftelie}}, \bibinfo {author} {\bibfnamefont {Connor}\ \bibnamefont {Powers}}, \ and\ \bibinfo {author} {\bibfnamefont {Wibe~A}\ \bibnamefont {De~Jong}},\ }\bibfield  {title} {\enquote {\bibinfo {title} {Arqtic: A full-stack software package for simulating materials on quantum computers},}\ }\href@noop {} {\bibfield  {journal} {\bibinfo  {journal} {ACM Transactions on Quantum Computing}\ }\textbf {\bibinfo {volume} {3}},\ \bibinfo {pages} {1--17} (\bibinfo {year} {2022}{\natexlab{c}})}\BibitemShut {NoStop}%
\bibitem [{\citenamefont {Kökcü}\ \emph {et~al.}(2022)\citenamefont {Kökcü}, \citenamefont {Steckmann}, \citenamefont {Wang}, \citenamefont {Freericks}, \citenamefont {Dumitrescu},\ and\ \citenamefont {Kemper}}]{K_kc__2022}%
  \BibitemOpen
  \bibfield  {author} {\bibinfo {author} {\bibfnamefont {Efekan}\ \bibnamefont {Kökcü}}, \bibinfo {author} {\bibfnamefont {Thomas}\ \bibnamefont {Steckmann}}, \bibinfo {author} {\bibfnamefont {Yan}\ \bibnamefont {Wang}}, \bibinfo {author} {\bibfnamefont {J.~K.}\ \bibnamefont {Freericks}}, \bibinfo {author} {\bibfnamefont {Eugene~F.}\ \bibnamefont {Dumitrescu}}, \ and\ \bibinfo {author} {\bibfnamefont {Alexander~F.}\ \bibnamefont {Kemper}},\ }\bibfield  {title} {\enquote {\bibinfo {title} {Fixed depth hamiltonian simulation via cartan decomposition},}\ }\href {\doibase 10.1103/physrevlett.129.070501} {\bibfield  {journal} {\bibinfo  {journal} {Physical Review Letters}\ }\textbf {\bibinfo {volume} {129}} (\bibinfo {year} {2022}),\ 10.1103/physrevlett.129.070501}\BibitemShut {NoStop}%
\bibitem [{\citenamefont {Trotter}(1959{\natexlab{b}})}]{Trotter1959}%
  \BibitemOpen
  \bibfield  {author} {\bibinfo {author} {\bibfnamefont {H.~F.}\ \bibnamefont {Trotter}},\ }\bibfield  {title} {\enquote {\bibinfo {title} {On the product of semi-groups of operators},}\ }\href@noop {} {\bibfield  {journal} {\bibinfo  {journal} {Proc. Am. Math. Soc.}\ }\textbf {\bibinfo {volume} {10}},\ \bibinfo {pages} {545--551} (\bibinfo {year} {1959}{\natexlab{b}})}\BibitemShut {NoStop}%
\end{thebibliography}%

\clearpage

\appendix
\section{Majorana Fermions in the TFIM}

In this section, we derive the Majorana fermion representation of the Transverse Field Ising Model. We begin with the TFIM Hamiltonian for $N+1$ sites,
\begin{equation}
    H = J \sum_{j=1}^{N} \sigma^x_j \sigma^x_{j+1} + h \sum_{j=1}^{N+1} \sigma^z_j.
    \label{eq:tfim}
\end{equation}

To map spin operators to fermionic operators, we apply the Jordan-Wigner transformation,
\begin{align}
    \sigma^z_j &= c^\dagger_j c_j -\frac{1}{2}\\
    \sigma^x_j &= \left( \prod_{k=1}^{j-1} (1 - 2 c_k^\dagger c_k) \right) c_j^\dagger +\left( \prod_{k=1}^{j-1} (1 - 2 c_k c_k^\dagger) \right)c_j
\end{align}

With this transformation, the TFIM Hamiltonian becomes a quadratic fermionic Hamiltonian,
\begin{equation}
    H = J \sum_{j=1}^{N} (c_j^\dagger - c_j)(c_{j+1}^\dagger + c_{j+1}) + h \sum_{j=1}^{N+1} ( c_j^\dagger c_j - \frac{1}{2})
\end{equation}

Now we introduce Majorana fermions:
\begin{equation}
    \gamma_{2j-1} = c_j + c_j^\dagger, \quad \gamma_{2j} = i(c_j - c_j^\dagger)
\end{equation}
These operators satisfy the Majorana conditions $\gamma_n^\dagger = \gamma_n$ and $\{\gamma_m, \gamma_n\} = 2\delta_{mn}$.

In terms of these Majorana operators, the TFIM Hamiltonian becomes
\begin{equation}
    H =  iJ \sum_{j=1}^{N} \gamma_{2j} \gamma_{2j+1} + ih \sum_{j=1}^{N+1} \gamma_{2j-1} \gamma_{2j}
    \label{eq:majorana}
\end{equation}
This is Eq. 2, which shows the TFIM as a quadratic Hamiltonian in Majorana fermions, clearly exposing the edge modes and topological structure of the chain in the limit of vanishing or dominant transverse field.

\section{Details on the Heatmap Near the Transition Regime}

As discussed in Section~IV.B, we identified two distinct dynamical regimes by varying the relative strengths of \( J \) and \( h \), and visualized this behavior using a heatmap. In this appendix, we provide additional site-resolved magnetization data to offer a more detailed view of the crossover region, particularly when \( J \approx h \).

These supplementary plots focus on values of \( J \) and \( h \) near the transition point to illustrate how the system gradually evolves from one regime to the other. This helps clarify the nature of the crossover and provides further evidence of how the interplay between intra-cell and inter-cell couplings governs the propagation and localization of excitations.

\begin{figure}
\subcaptionbox{\protect\label{a}}{\includegraphics[width=0.48\linewidth,height=0.42\linewidth]{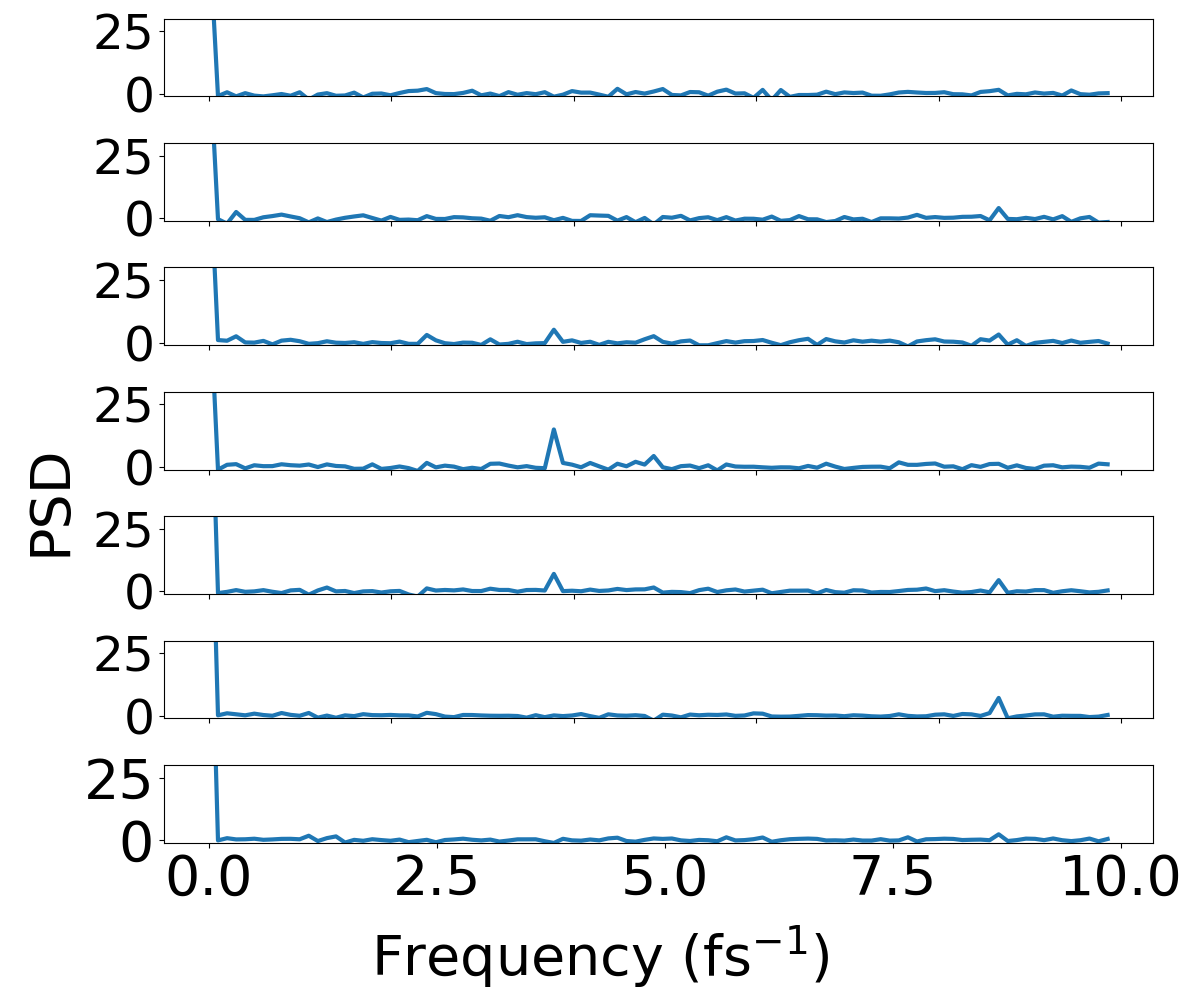}}\hfill
\subcaptionbox{\protect\label{b}}
{\includegraphics[width=0.48\linewidth,height=0.42\linewidth]{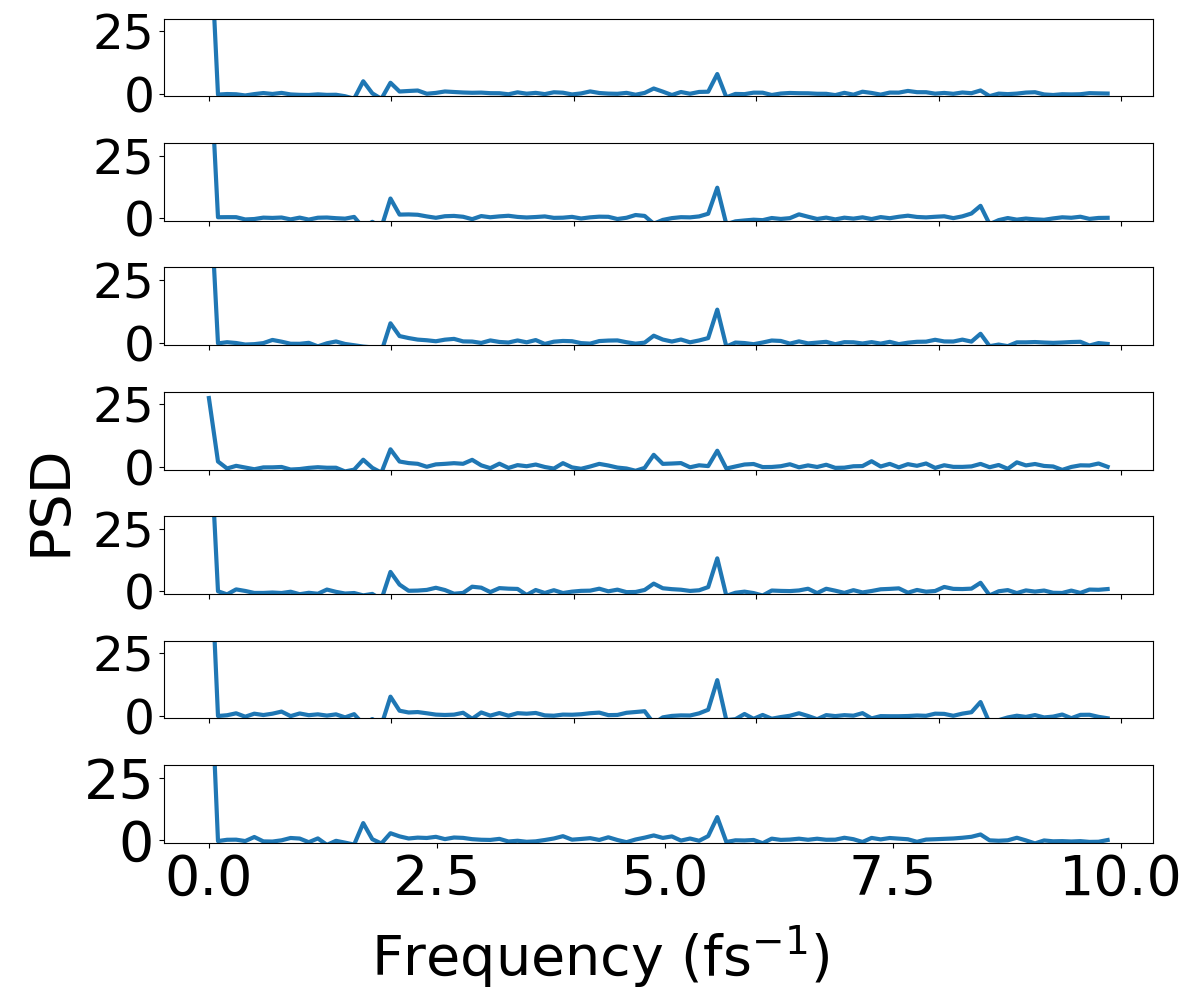}}\hfill
\subcaptionbox{\protect\label{c}}
{\includegraphics[width=0.48\linewidth,height=0.42\linewidth]{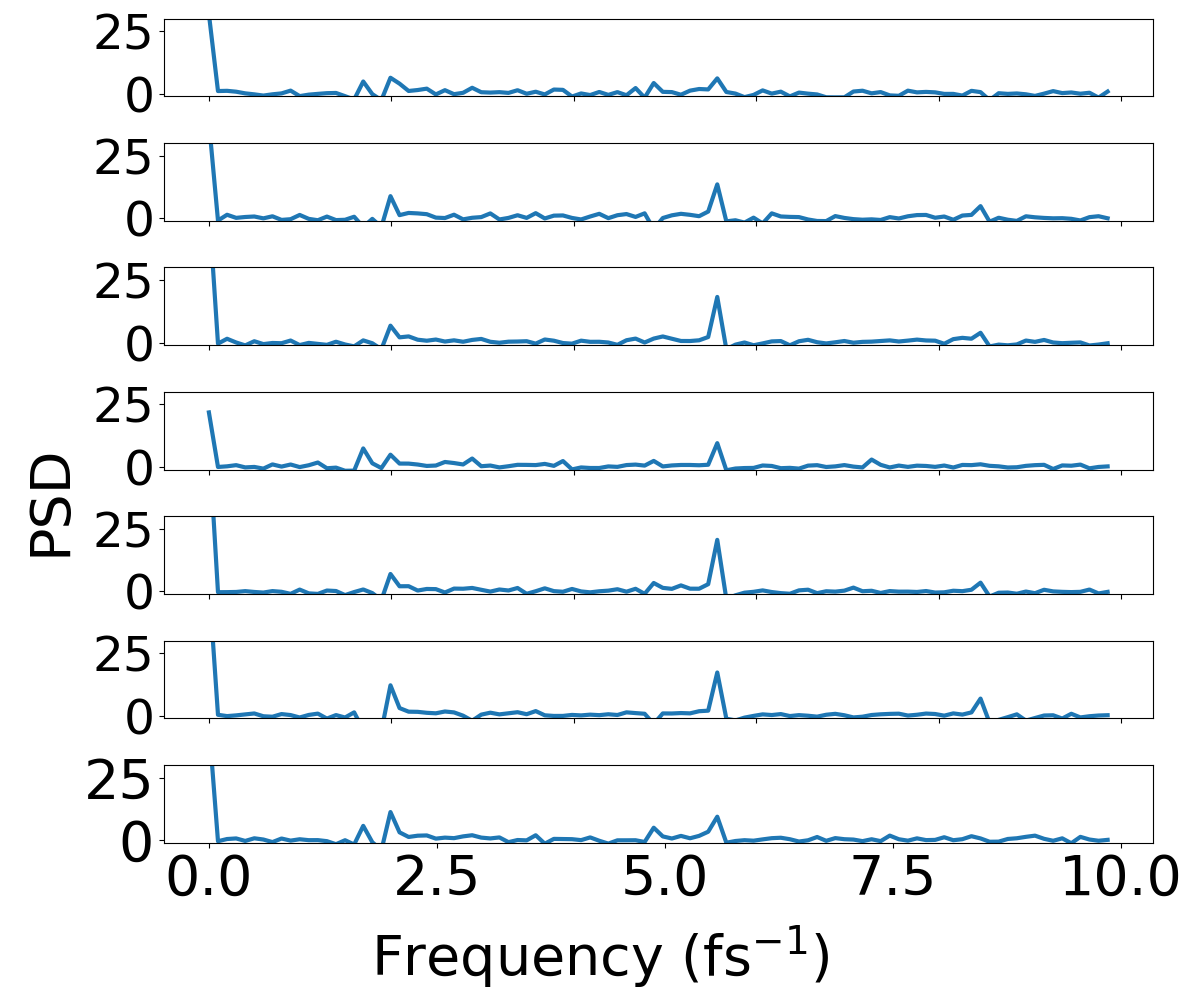}}\hfill
\subcaptionbox{\protect\label{d}}
{\includegraphics[width=0.48\linewidth,height=0.42\linewidth]{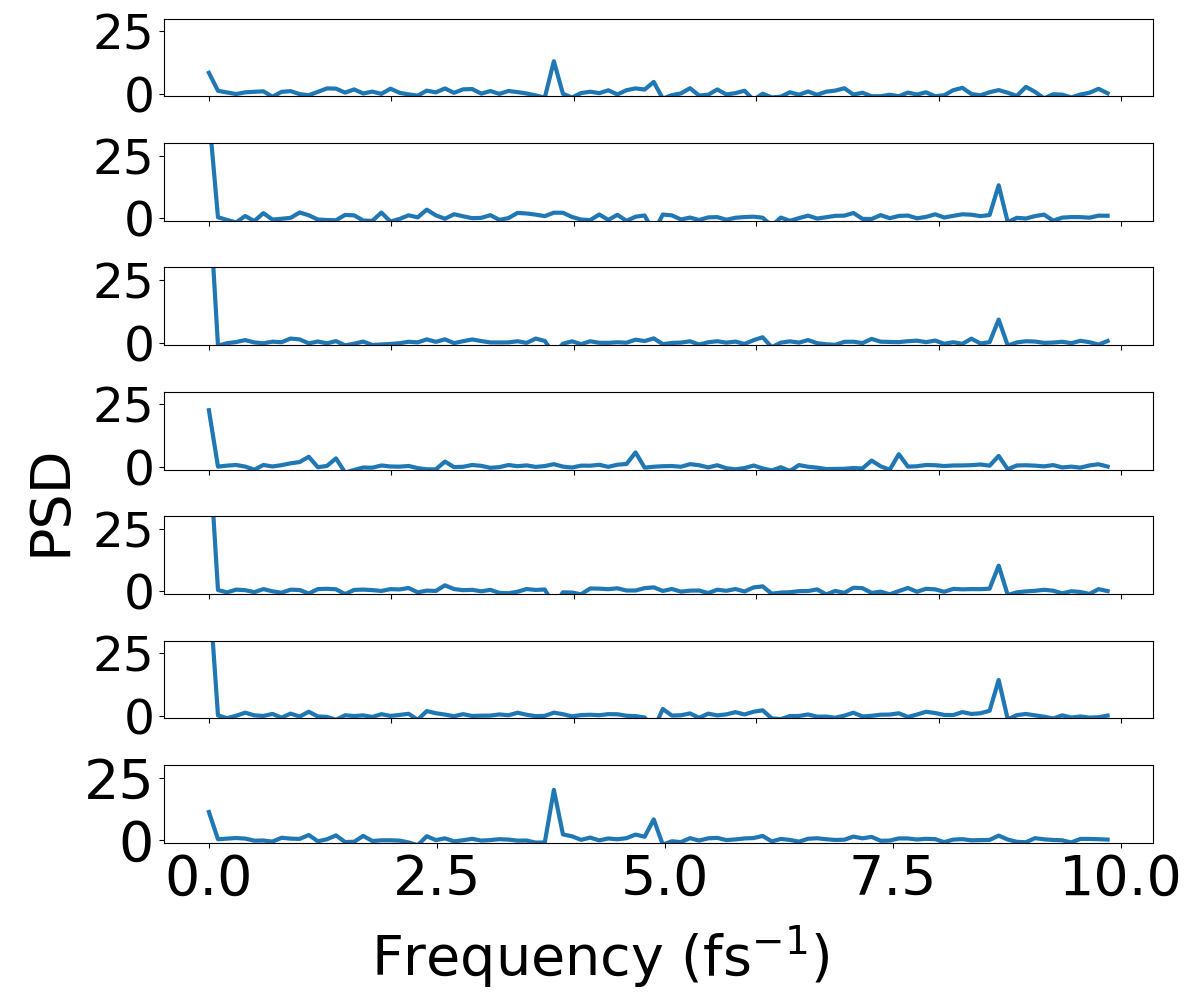}}\hfill
\caption{Fourier transform of the individual qubit magnetization in the $z$-direction for varying coupling strengths and magnetic fields, with fixed $\lambda = 1.0$. (a) $J = 4.0$~eV, $h = 7.0$~eV; (b) $J = 5.0$~eV, $h = 6.0$; (c) $J = 6.0$~eV, $h = 5.0$~eV; (d) $J = 7.0$~eV, $h = 4.0$~eV.}
\end{figure}

\section{Derivation of circuit}\label{Appendix:circuit derivation}
\begin{figure}
    \centering
    \includegraphics[width=1.0\linewidth,height=0.3\linewidth]{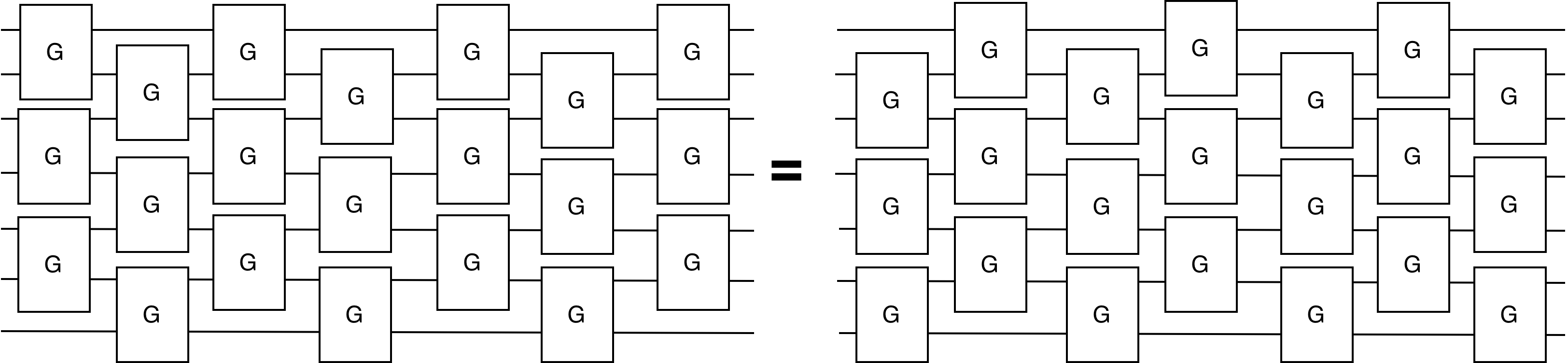}
    \caption{The mirroring symmetry in a 7-qubit circuit.}
    \label{figure:mirroring}
\end{figure}

\begin{figure}
\captionsetup[subfigure]{position=bottom}
    \centering
    \subcaptionbox{\protect\label{a}}{\includegraphics[width=0.48\linewidth,height=0.20\linewidth]{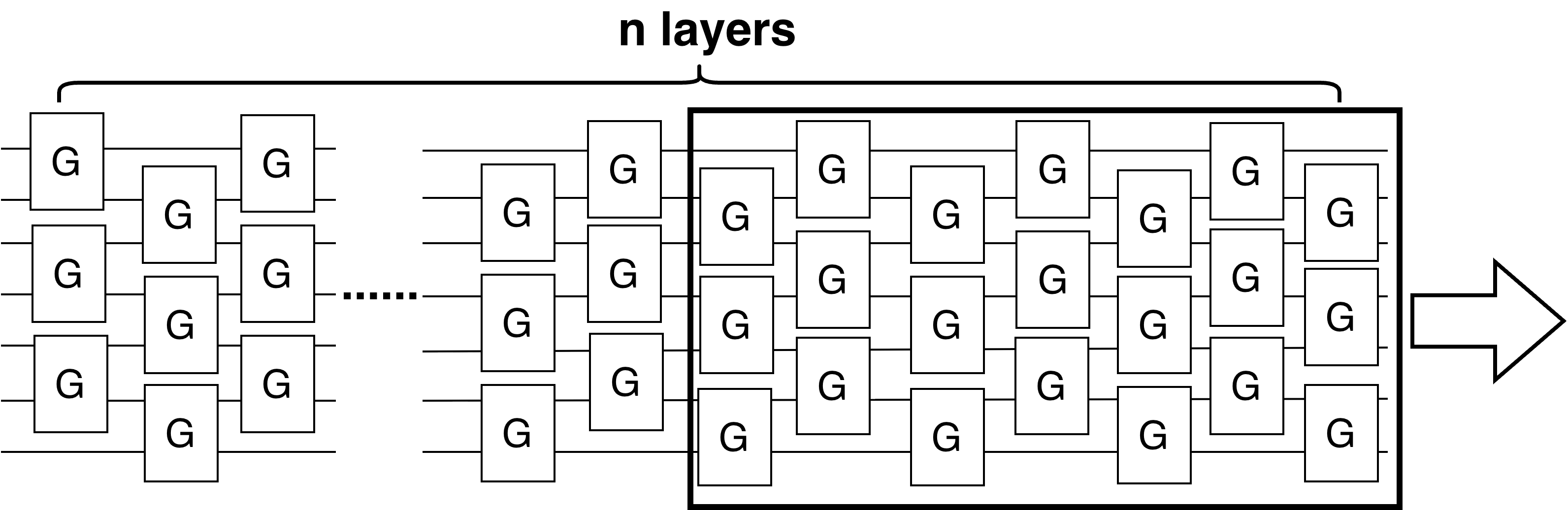}}\hfill
    \subcaptionbox{\protect\label{b}}{\includegraphics[width=0.48\linewidth,height=0.20\linewidth]{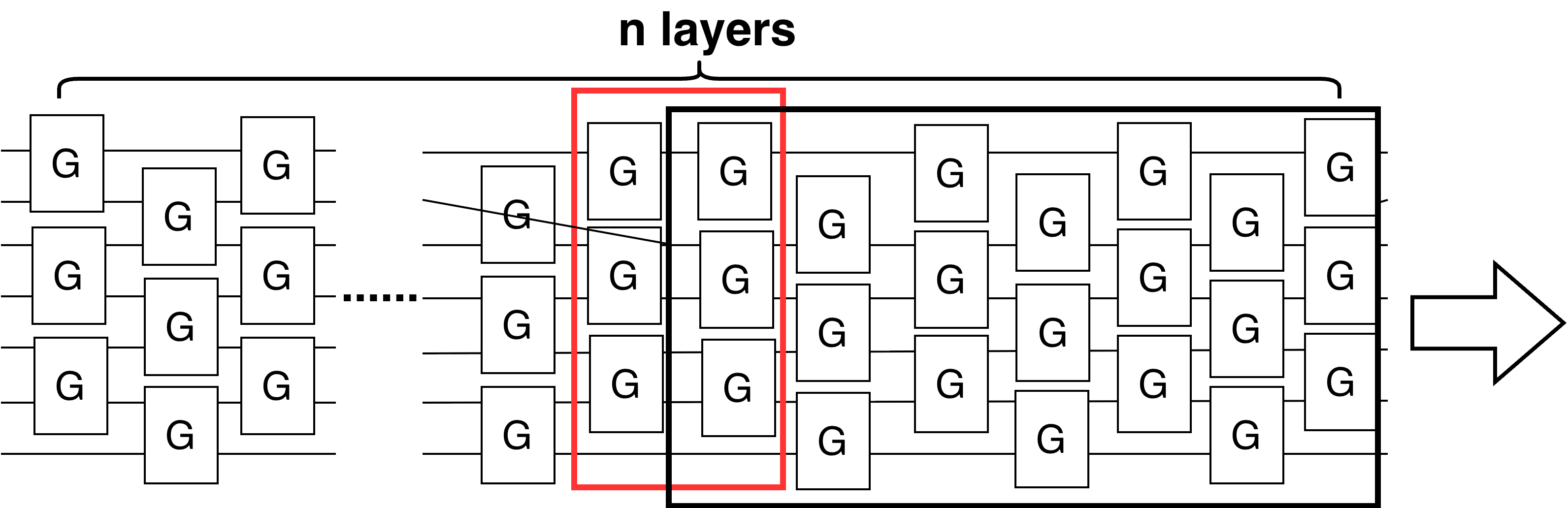}}\hfill
    \subcaptionbox{\protect\label{c}}{\includegraphics[width=0.63\linewidth,height=0.20\linewidth]{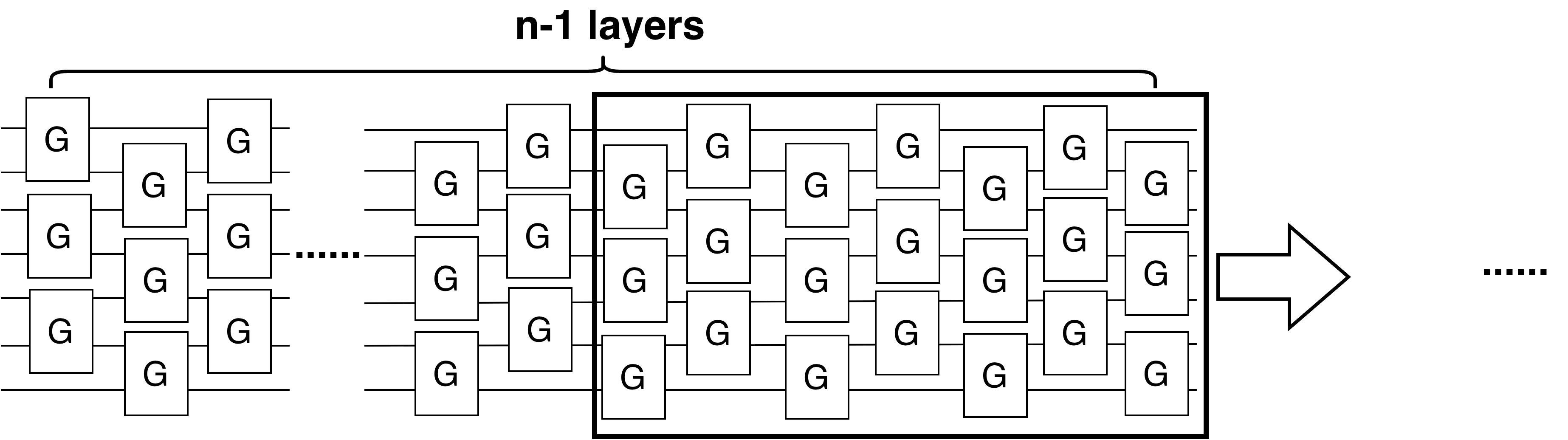}}\hfill
    \subcaptionbox{\protect\label{d}}{\includegraphics[width=0.35\linewidth,height=0.20\linewidth]{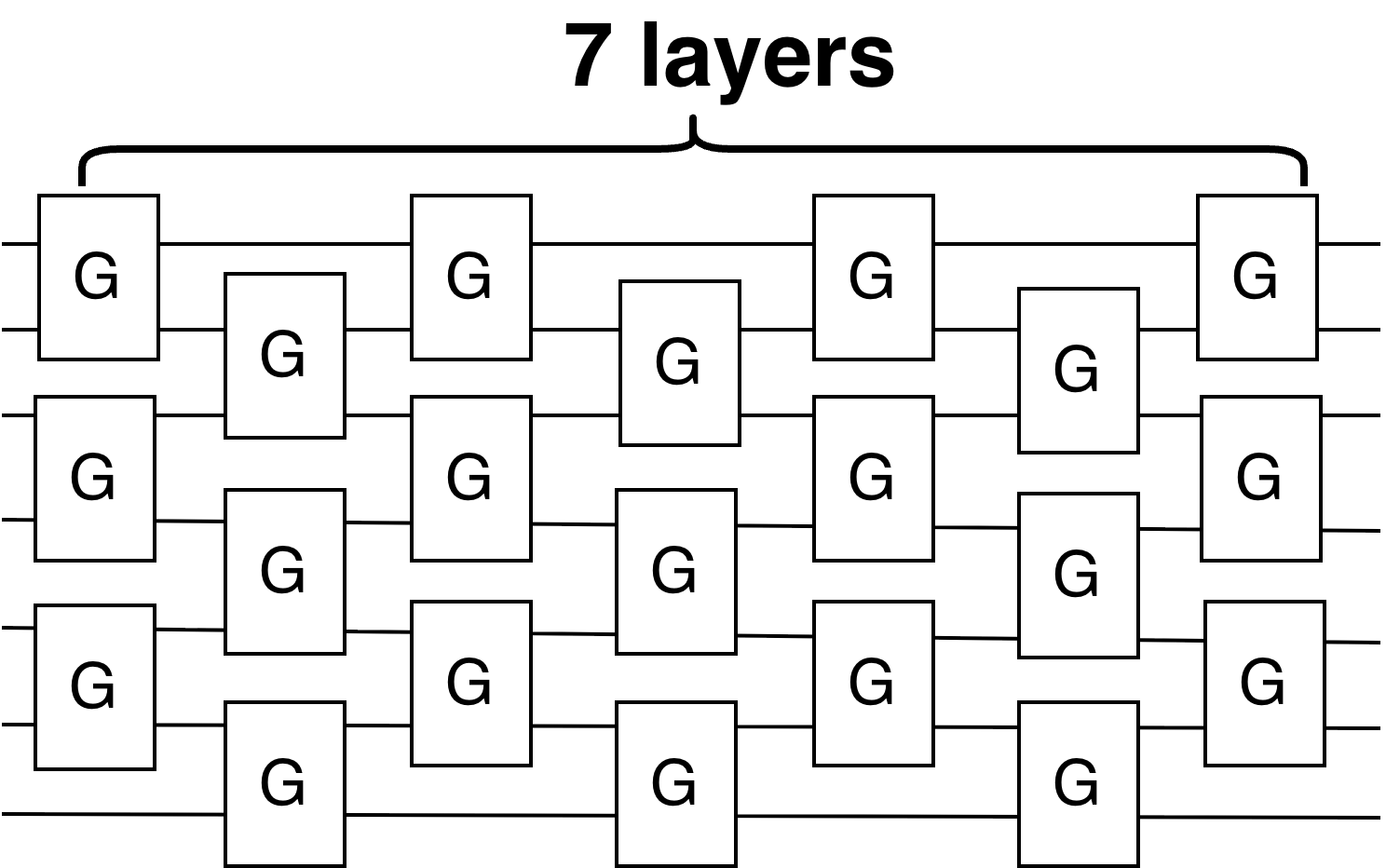}}\hfill
    \caption{Derivation of the constant-depth quantum circuit leveraging the mirroring symmetry of the system and the composability of sequential gate operations. (a) The construction of the circuit for depth $n$ is obtained by unitary time evolution using Trotter decomposition \cite{Trotter1959}. 
    (b) The portion of circuit in the black rectangle is transformed according to mirror symmetry \cite{Bassman_MT2022}. 
    (c) Application of the property G$^2$=G to the red rectangle in panel (b) leads to the contraction of 1 layer, transforming the starting circuit of depth $n$ into this circuit of depth $n-1$. (d) Repetition of the cycle (a)$\rightarrow$(b) and (b)$\rightarrow$(c) a number of times equal to $n-7$ leads to the compact and constant-depth circuit with 7 layers. }
%
    \label{figure:Final derivation}
\end{figure}

With the property of Mirroring, we are able to mirror a 7-qubit circuit as shown in Fig.~\ref{figure:mirroring}. Moreover, for any quadratic Hamiltonian, we establish the model using the circuit in Fig.~\ref{figure:Final derivation}(a). The last seven qubits can be mirrored using the rule shown in Fig.~\ref{figure:mirroring}. Following the property of compatibility, the column shown in Fig.~\ref{figure:Final derivation}(b) are reduced to one. Then, we choose the last seven qubits as in Fig.~\ref{figure:Final derivation}(a). So on and so forth, the circuit can be reduced to only 7 qubits, shown in Fig.~\ref{figure:Final derivation}.

\end{document}